\documentclass[longauth]{aa} 
\usepackage{graphicx}
\usepackage{natbib}
\bibpunct{(}{)}{;}{a}{}{,} % to follow the A&A style
\usepackage{txfonts}
\usepackage{booktabs} % For better tables (optional)
\usepackage{xcolor}
\usepackage{float}
\usepackage{tabularx}
\definecolor{darkgreen}{rgb}{0.0, 0.5, 0.0}
\usepackage[colorlinks,citecolor=blue,linkcolor=red,hyperfootnotes=true,urlcolor=blue]{hyperref}
\usepackage{subfig} % Alternative to subcaption
\usepackage{longtable} % For multipage tables
\usepackage{booktabs}  % For nice table lines
\usepackage{array}     % For better column formatting
\usepackage{threeparttable}
\usepackage{comment}
\usepackage{orcidlink}
\usepackage[normalem]{ulem}
\usepackage[utf8]{inputenc}

\defcitealias{AlmenaraBonfils2022}{A22}
\defcitealias{Osborne2024}{O24}
\defcitealias{Delisle2022}{D22}
\defcitealias{Stefanov2025}{S25}
\defcitealias{Ahrer2025}{A25}

\newcommand{\Pcyc}{$3370 \pm 120$}
\newcommand{\Prot}{$17.95 \pm 0.15$}
\newcommand{\massb}{$4.52 \pm 0.47$}
\newcommand{\massc}{$10.0 \pm 1.3$}
\newcommand{\masscandidateone}{$4.15 \pm 1.04$}
\newcommand{\bestkernel}{MEP }
\newcommand{\inclination}{$86.93 \pm 0.17$}
\newcommand{\period}{$2.85310198 \pm 8.2\times 10^{-7}$}
\newcommand{\tc}{$1356.15322 \pm 0.00023$}
\newcommand{\density}{$2.40^{+0.33}_{-0.30}$}
\newcommand{\radius}{$2.18 \pm 0.06$}

\newcommand{\teff}{\ensuremath{T_{\rm eff}}}
\newcommand{\feh}{\mbox{[Fe/H]}\xspace}

\DeclareMathOperator*{\best}{best}

\begin{document} 
% Define variables after document starts
    
    \title{NIRPS tightens the mass estimate of GJ~3090\,b and detects a planet near the stellar rotation period}
    
    \titlerunning{The GJ~3090 system}
    \authorrunning{P. Lamontagne, et al.}
    
    \author{
Pierrot Lamontagne\inst{1,*},
Drew Weisserman\inst{2}\orcidlink{0000-0002-7992-469X},
Charles Cadieux\inst{1}\orcidlink{0000-0001-9291-5555},
David Lafreni\`ere\inst{1}\orcidlink{0000-0002-6780-4252},
Alexandrine L'Heureux\inst{1}\orcidlink{0009-0005-6135-6769},
Mykhaylo Plotnykov\inst{3}\orcidlink{0000-0002-9479-2744},
L\'ena Parc\inst{4}\orcidlink{0000-0002-7382-1913},
Atanas K. Stefanov\inst{5,6}\orcidlink{0000-0002-6059-1178},
Leslie Moranta\inst{1,7}\orcidlink{0000-0001-7171-5538},
Ren\'e Doyon\inst{1,8}\orcidlink{0000-0001-5485-4675}
% ...
}

\institute{
\inst{1}Institut Trottier de recherche sur les exoplan\`etes, D\'epartement de Physique, Universit\'e de Montr\'eal, Montr\'eal, Qu\'ebec, Canada\\
\inst{2}Department of Physics \& Astronomy, McMaster University, 1280 Main St W, Hamilton, ON, L8S 4L8, Canada\\
\inst{3}Department of Physics, University of Toronto, Toronto, ON M5S 3H4, Canada\\
\inst{4}Observatoire de Gen\`eve, D\'epartement d’Astronomie, Universit\'e de Gen\`eve, Chemin Pegasi 51, 1290 Versoix, Switzerland\\
\inst{5}Instituto de Astrof\'isica de Canarias (IAC), Calle V\'ia L\'actea s/n, 38205 La Laguna, Tenerife, Spain\\
\inst{6}Departamento de Astrof\'isica, Universidad de La Laguna (ULL), 38206 La Laguna, Tenerife, Spain\\
\inst{7}Plan\'etarium de Montr\'eal, Espace pour la Vie, 4801 av. Pierre-de Coubertin, Montr\'eal, Qu\'ebec, Canada\\
\inst{8}Observatoire du Mont-M\'egantic, Qu\'ebec, Canada\\
\inst{9}European Southern Observatory (ESO), Karl-Schwarzschild-Str. 2, 85748 Garching bei M\"unchen, Germany\\
\inst{10}University Observatory, Faculty of Physics, Ludwig-Maximilians-Universit\"at M\"unchen, Scheinerstr. 1, 81679 Munich, Germany\\
\inst{11}European Southern Observatory (ESO), Av. Alonso de Cordova 3107,  Casilla 19001, Santiago de Chile, Chile\\
\inst{12}Instituto de Astrof\'isica e Ci\^encias do Espa\c{c}o, Universidade do Porto, CAUP, Rua das Estrelas, 4150-762 Porto, Portugal\\
\inst{13}Departamento de F\'isica e Astronomia, Faculdade de Ci\^encias, Universidade do Porto, Rua do Campo Alegre, 4169-007 Porto, Portugal\\
\inst{14}Department of Earth, Planetary, and Space Sciences, University of California, Los Angeles, CA 90095, USA\\
\inst{15}Univ. Grenoble Alpes, CNRS, IPAG, F-38000 Grenoble, France\\
\inst{16}Departamento de F\'isica Te\'orica e Experimental, Universidade Federal do Rio Grande do Norte, Campus Universit\'ario, Natal, RN, 59072-970, Brazil\\
\inst{17}Department of Physics, McGill University, 3600 rue University, Montr\'eal, QC, H3A 2T8, Canada\\
\inst{18}Department of Earth \& Planetary Sciences, McGill University, 3450 rue University, Montr\'eal, QC, H3A 0E8, Canada\\
\inst{19}Departamento de F\'isica, Universidade Federal do Cear\'a, Caixa Postal 6030, Campus do Pici, Fortaleza, Brazil\\
\inst{20}Centro de Astrobiolog\'ia (CAB), CSIC-INTA, Camino Bajo del Castillo s/n, 28692, Villanueva de la Ca\~nada (Madrid), Spain\\
\inst{21}Centre Vie dans l’Univers, Facult\'e des sciences de l’Universit\'e de Gen\`eve, Quai Ernest-Ansermet 30, 1205 Geneva, Switzerland\\
\inst{22}Space Research and Planetary Sciences, Physics Institute, University of Bern, Gesellschaftsstrasse 6, 3012 Bern, Switzerland\\
\inst{23}Consejo Superior de Investigaciones Cient\'ificas (CSIC), E-28006 Madrid, Spain\\
\inst{24}Bishop's Univeristy, Dept of Physics and Astronomy, Johnson-104E, 2600 College Street, Sherbrooke, QC, Canada, J1M 1Z7, Canada\\
\inst{25}Department of Physics, Engineering Physics, and Astronomy, Queen’s University, 99 University Avenue, Kingston, ON K7L 3N6, Canada\\
\inst{26}Department of Physics and Space Science, Royal Military College of Canada, 13 General Crerar Cres., Kingston, ON K7P 2M3, Canada\\
\inst{27}Instituto de Astrof\'isica e Ci\^encias do Espa\c{c}o, Faculdade de Ci\^encias da Universidade de Lisboa, Campo Grande, 1749-016 Lisboa, Portugal\\
\inst{28}Departamento de F\'isica da Faculdade de Ci\^encias da Universidade de Lisboa, Edif\'icio C8, 1749-016 Lisboa, Portugal\\
\inst{29}Centre of Optics, Photonics and Lasers, Universit\'e Laval, Qu\'ebec, Canada\\
\inst{30}Herzberg Astronomy and Astrophysics Research Centre, National Research Council of Canada, Canada\\
\inst{31}Aix Marseille Univ, CNRS, CNES, LAM, Marseille, France\\
\inst{32}Center for Space and Habitability, University of Bern, Gesellschaftsstrasse 6, 3012 Bern, Switzerland\\
\inst{33}Center for astrophysics $\vert$ Harvard \& Smithsonian, 60 Garden Street, Cambridge, MA 02138, USA\\
\inst{34}NASA Exoplanet Science Institute, IPAC, California Institute of Technology, Pasadena, CA 91125, USA\\
\inst{35}George Mason University, 4400 University Drive, Fairfax, VA, 22030, USA\\
\inst{36}Division of Astrophysics, Department of Physics, Lund University, Box 118, SE-22100 Lund, Sweden\\
\inst{37}SETI Institute, Mountain View, CA 94043, USA NASA Ames Research Center, Moffett Field, CA 94035, USA\\
\inst{38}York University, 4700 Keele St, North York, ON M3J 1P3, Canada\\
\inst{39}Max-Planck-Institut f\"ur Astronomie, K\"onigstuhl 17, D-69117 Heidelberg, Germany\\
\inst{40}University of British Columbia, 2329 West Mall, Vancouver, BC, Canada, V6T 1Z4, Canada\\
\inst{41}Western University, Department of Physics \& Astronomy and Institute for Earth and Space Exploration, 1151 Richmond Street, London, ON N6A 3K7, Canada\\
\inst{42}Light Bridges S.L., Observatorio del Teide, Carretera del Observatorio, s/n Guimar, 38500, Tenerife, Canarias, Spain\\
\inst{43}Department of Physics, The University of Warwick, Gibbet Hill Road, Coventry, CV4 7AL, UK\\
\inst{44}Hamburger Sternwarte, Gojenbergsweg 112, D-21029 Hamburg, Germany\\
\inst{45}Subaru Telescope, National Astronomical Observatory of Japan (NAOJ), 650 N Aohoku Place, Hilo, HI 96720, USA\\
\inst{46}Department of Astronomy \& Astrophysics, University of Chicago, 5640 South Ellis Avenue, Chicago, IL 60637, USA\\
\inst{47}Laboratoire Lagrange, Observatoire de la C\^ote d’Azur, CNRS, Universit\'e C\^ote d’Azur, Nice, France\\
\inst{*}\email{pierrot.lamontagne@umontreal.ca}
}

\date{Received 16 August 2025 / Accepted 15 December 2025}

\abstract{
We present an updated characterization of the planetary system orbiting the nearby M2 dwarf GJ~3090 (TOI-177; $d = 22$~pc), based on new high-precision radial velocity (RV) observations from NIRPS and HARPS. With an orbital period of 2.85~d, the transiting sub-Neptune GJ~3090\,b has a mass we refine to \massb\,$M_{\oplus}$, which, combined with our derived radius of \radius\,$R_{\oplus}$, yields a density of \density\,g\,cm$^{-3}$. The combined interior structure and atmospheric constraints indicate that GJ~3090\,b is a compelling water-world candidate, with a volatile-rich envelope in which water likely represents a significant fraction. We also confirm the presence of a second planet, GJ~3090\,c, a sub-Neptune with a 15.9~d orbit and a minimum mass of \massc\,$M_{\oplus}$, which does not transit. Despite its proximity to the star’s 18~d rotation period, our joint analysis using a multidimensional Gaussian process (GP) model that incorporates TESS photometry and differential stellar temperature measurements distinguishes this planetary signal from activity-induced variability. In addition, we place new constraints on a non-transiting planet candidate with a period of 12.7~d, suggested in earlier RV analyses. This candidate remains a compelling target for future monitoring. These results highlight the crucial role of multidimensional GP modelling in disentangling planetary signals from stellar activity, enabling the detection of a planet near the stellar rotation period that could have remained undetected with traditional approaches.
}

\maketitle
%
%-------------------------------------------------------------------
\clearpage
\section{Introduction}\label{sec:intro}
M dwarf stars offer a favourable environment for exoplanet studies. Comprising approximately 70\% of the stars in our galaxy \citep[e.g.][]{Reid1997, Bochanski2010} and representing the majority of nearby stellar systems \citep[e.g.][]{Henry2006}, these  low-mass low-luminosity stars amplify detectable signals through both radial velocity and transit methods. Their smaller radius makes transit detection more feasible, as planets occult a larger fraction of the stellar disk, while their lower mass enhances the gravitational reflex motion induced by orbiting planets. Additionally, M dwarfs have a higher occurrence rate of small planets \citep[e.g.][]{Bonfils2013, Sabotta2021}, and this combined with the enhanced detectability of atmospheric features means they are ideal targets for studying their composition and evolutionary pathways \citep[e.g.][]{Benneke2019, Kreidberg2019}.

Sub-Neptunes stand out as the most abundant exoplanets in our galaxy \citep{Bonfils2013,Dressing2015,Fulton_2017}, yet their nature and formation mechanisms remain poorly understood. With radii between those of Earth and Neptune, they occupy a focus of parameter space where composition can vary considerably. This diversity is illustrated by the radius valley \citep{Fulton_2017,Cloutier2020}, a decrease in the number of observed planets between 1.5 and 2.0 $R_{\oplus}$, suggesting distinct formation and evolution mechanisms that shape these planets' structure and composition. Various scenarios have been proposed to explain this gap, including atmospheric loss through photoevaporation \citep{Owen2017}, core-powered mass loss \citep{Gupta2020}, intrinsically gas-poor formation \citep{Lee2016,Venturini2020}, or compositions dominated by volatiles \citep{Zeng2019,Nixon2021,Aguichine2021}.

GJ~3090\,b, first identified by \citet{AlmenaraBonfils2022} (hereafter \citetalias{AlmenaraBonfils2022}) exemplifies the importance of the research on sub-Neptunes. With a reported radius of $2.171 \pm 0.068 R_{\oplus}$ \citep[hereafter \citetalias{Ahrer2025}]{Ahrer2025}, GJ~3090\,b lies near the transition between super-Earths and sub-Neptunes, providing a unique opportunity to study atmospheric retention and interior composition \citep{Rogers2015}. Sub-Neptunes in this regime exhibit a wide diversity in bulk densities, ranging from rocky cores with thin gaseous envelopes to volatile-rich water worlds \citep{Zeng2019}. Radial velocity (RV) measurements are crucial for constraining the masses of such planets, which in turn enable bulk density calculations and interior structure modelling \citep{Borsato2019, Hoyer2021, Leleu2024}. GJ~3090\,b's initial mass estimate of $3.34 \pm 0.72$\,M$_\oplus$ and equilibrium temperature of $693 \pm 18$\,K, derived from High Accuracy Radial velocity Planet Searcher \citep[HARPS;][]{MayorPepe2003} measurements, placed it among the lower-density sub-Neptunes. However, precise mass constraints are critical for distinguishing between competing formation and evolutionary scenarios \citep{Otegi2020}.

Advances in instrumentation, including the new Near InfraRed Planet Searcher \citep[NIRPS;][]{Bouchy2025} at La Silla Observatory, have enhanced the precision of RV measurements in the near-infrared. NIRPS and HARPS have the capability to observe simultaneously, which can be useful for identifying stellar activity in RV data using its chromatic nature. In addition, new RV analysis techniques, including multidimensional Gaussian processes \citep[GPs;][]{Rajpaul2015}, are now able to accurately separate planetary and stellar contributions in radial velocities, which enables us to precisely characterize known planets and find additional companions. As part of the NIRPS Guaranteed Time Observations (GTO) programme, GJ~3090 was monitored with the aim of improving the mass determination of its transiting planet GJ~3090\,b. By combining these new NIRPS observations with an expanded HARPS dataset, which  resulted in around 150 additional radial velocity measurements, we present a comprehensive reanalysis of the GJ~3090 system.

Section~\ref{sec:observations} details the observational datasets used in this work, including photometry, RVs, and differential temperatures ($\Delta T$). In Section~\ref{sec:stellar_parameters} we characterize the stellar properties of GJ~3090. Then we present the analysis of the photometry datasets in Section \ref{sect:photometry_analysis}, from which we update the radius of GJ~3090\,b and identify evidence for a possible long-term variability signal that may be associated with a magnetic cycle. Our RV modelling framework is introduced in Section~\ref{sec:modelling}, where we describe the treatment of instrumental offsets and jitter, the correction for this long-term signal, the modelling of Keplerian planetary signals, and the implementation of a multidimensional GP to mitigate stellar activity. The results are presented in Section~\ref{sec:results}, including updated mass and density constraints for GJ~3090\,b and a search for additional planetary signatures in the radial velocities and photometry. Finally, we discuss the architecture of the GJ~3090 system in a broader exoplanetary context. We conclude with a summary of our findings in Section~\ref{sec:conclusion}.

\section{Observations} \label{sec:observations}
\subsection{TESS photometry} \label{sect:tess_juliet}
GJ~3090 (TIC~262530407, TOI-177) was observed with TESS in four different sectors throughout its primary and extended missions: Sector 2 (2018-08-23 to 2018-09-20), Sector 3 (2018-09-20 to 2018-10-17), Sector 29 (2020-08-26 to 2020-09-21), and Sector 69 (2023-08-25 to 2023-09-20). For our study, we used both the Simple Aperture Photometry \citep[SAP;][]{Twicken2010,Morris2020} and the Presearch Data Conditioning SAP \citep[PDCSAP;][]{Stumpe2012,Stumpe2014,Smith2012} data products provided by the TESS Science Processing Operations Center \citep[SPOC;][]{Jenkins2016} at NASA Ames Research Center, available through the Mikulski Archive for Space Telescopes.\footnote{\href{https://archive.stsci.edu/missions-and-data/tess}{archive.stsci.edu/tess/}} On the one hand, as in \citetalias{AlmenaraBonfils2022}, we chose the SAP photometry over the PDCSAP data to study the stellar rotation (Section \ref{sec:modelling}), as the latter removes long-term trends and, in this case, was found to overcorrect this signal. For our stellar activity analysis, we removed in-transit photometric data points from the light curve to isolate the out-of-transit flux variations. This step was essential for using the light curve as a stellar activity indicator in our RV analysis. To reduce computation time while retaining adequate sampling to model the 18 d stellar rotation period found by \citetalias{AlmenaraBonfils2022}, we binned the TESS light curves into 8-hour intervals.

Since TESS collected an additional sector (Sector 69) after the publication of \citetalias{AlmenaraBonfils2022}, we updated the transit analysis accordingly, using the PDCSAP flux, which accounts for dilution caused by contaminating sources within the TESS aperture. This analysis is described in Sect.~\ref{sect:tess_juliet}.

\subsection{ASAS-SN photometry} \label{subsec:ASASN-photometry}
To investigate the long-term variability of GJ~3090 and assess the possible presence of a magnetic cycle, we used archival photometric data from the All-Sky Automated Survey for Supernovae (ASAS-SN). ASAS-SN is a global network of 14\,cm telescopes designed to monitor the entire visible sky on a nightly basis, originally developed to detect supernovae and other transient phenomena \citep{Kochanek2017}. The network consists of 24 telescopes distributed across six observatories worldwide, ensuring consistent coverage and minimizing weather-related data gaps. Sites are located at: Haleakal$\bar{\mathrm{a}}$ Observatory, HI, USA; two at Cerro Tololo Observatory, Coquimbo, Chile; South African Astronomical Observatory, Sutherland, South Africa; McDonald Observatory, TX,
USA; Ali Observatory, Tibet, China. 

The ASAS-SN light curve was retrieved from the Sky Patrol V1 database,\footnote{\href{https://asas-sn.osu.edu/}{https://asas-sn.osu.edu/}} with a manual correction for the stellar proper motion applied by supplying the proper motion-corrected coordinates from \texttt{SIMBAD} in the query, and all measurements obtained when the Moon was within $90^\circ$ of the target were excluded. After a $3\sigma$ outlier removal, the dataset for GJ~3090 contains 2772 photometric measurements spanning May 2014 to September 2025, for a total time span of 4144 days (11.3 years). The data come from four different cameras labelled \textit{bf}, \textit{bj}, \textit{bF}, and \textit{bn}. The \textit{bf} camera is distinct from the others in that it provides the only data obtained before 2018, the year it became out of service, and it uses a slightly redder effective bandpass ($\lambda_\mathrm{eff} = 551$\,nm, $V$ filter) compared to the other three cameras ($\lambda_\mathrm{eff} = 480$\,nm, $g$ filter). The sampling is irregular, with a mean time interval between measurements of about four days. In this work, the ASAS-SN data are used to identify the presence of a long-term signal on GJ~3090, which can motivate its modelling in other time series. This is done as an independent analysis described in Section~\ref{cycle discussion}. 

\subsection{NIRPS radial velocities} \label{subsubsec:nirps-obs}

We acquired spectra of GJ~3090 on 98 nights with the NIRPS \citep{BouchyDoyon2017,WildiBouchy2022}, between 2022 Nov.\ 29 and 2025 Jan 10, as part of the NIRPS-GTO. NIRPS is an echelle spectrograph at the ESO 3.6-m telescope of the La Silla Observatory in Chile, with a wavelength range of $980-1800$\,nm. The instrument is equipped with a low-order adaptive optics system and two observing modes, High Accuracy ($R \sim 85000$, $0.4"$ fibre) and High Efficiency ($R \sim 70000$, $0.9"$ fibre) that can be utilized simultaneously with HARPS. All GJ~3090 observations used here are in the high-efficiency mode. Most data points were obtained by binning two 900s exposures done in the same night. We excluded from our analysis 14 night-binned spectra of shorter 300 s exposures acquired either during the NIRPS commissioning phase or for transit monitoring in another work package, as they exhibit higher dispersion and lower S/N than expected from photo-noise scaling. This conservative selection ensures a homogeneous dataset with consistent noise properties. The observations were reduced with the data reduction software \texttt{APERO} v0.7.292 \citep{CookArtigau2022}, fully compatible with NIRPS, with RV extraction performed using the line-by-line method implemented in the open-source \texttt{lbl} package,\footnote{\href{https://lbl.exoplanets.ca/}{lbl.exoplanets.ca/}} as described in \citet{ArtigauCadieux2022}. At the end, we are left with 84 data points and a median RV precision of $1.8$ m\,s$^{-1}$ per point. We built a template spectrum of GJ~3090 using the telluric-corrected spectra from \texttt{APERO} in order to derive stellar parameters independently from the literature and to obtain the abundances of several elements with spectral features in the NIRPS spectral range. The top panel of figure \ref{fig:timeseries} shows the last 59 RV measurements of NIRPS.

\subsection{HARPS radial velocities} \label{subsubsec:harps-obs}

\citetalias{AlmenaraBonfils2022} obtained 55 radial velocities of GJ~3090 with HARPS in high-accuracy mode (HAM). We also obtained 26 new HAM measurements and 33 EGGS (high-efficiency mode) measurements simultaneous with NIRPS. All simultaneous observations with NIRPS were obtained with the same exposure times and binning procedure. The data were reduced with the HARPS DRS3.2.5 and radial velocities were extracted with version 0.65.003 of the \texttt{lbl} package. The template spectrum used to derive the radial velocities of the HAM mode was obtained by combining the archival and contemporaneous spectra in order to increase its S/N. HAM and EGGS were reduced with their own template because of the difference in their spectral resolution. We obtain 82 HAM points with a median RV precision of $1.9$ m/s and 33 EGGS points with a median RV precision of $1.5$ m/s for a total of 109 HARPS measurements. The top panel of figure \ref{fig:timeseries} shows the last 45 RV measurements of HARPS. 

\subsection{HARPS (HAM) differential stellar temperatures} \label{subsec:dTemp}

The $\Delta T$ time series was extracted from the HARPS (HAM) spectra following the method introduced in \cite{Artigau2024}. This activity indicator measures the temperature difference between photospheric regions contributing to line asymmetries and the surrounding quiet stellar surface, providing a direct proxy for stellar heterogeneities such as spots and plages. Unlike traditional activity indicators, for example the H$\alpha$ index or bisector span, $\Delta T$ is grounded in physical temperature variations with a sub-Kelvin-precision. The $\Delta T$ dataset contains 78 measurements extracted from the HAM spectra with a median precision of $0.4$K (see \ref{fig:timeseries}). Of all activity indicators available in \texttt{lbl}, it was the one with the strongest signal around the rotation period of GJ~3090 found in \citetalias{AlmenaraBonfils2022}. We did not use the differential temperatures from either NIRPS or HARPS (EGGS), as both showed a weak correlation with the RVs, making the added complexity unjustified. In particular, including the NIRPS indicator would have required an additional GP dimension.

\section{Spectroscopic stellar parameters} \label{sec:stellar_parameters}

The effective temperature and chemical abundances of GJ~3090 are derived from the NIRPS template spectrum following the methodology of \citet{jahandarComprehensiveHighresolutionChemical2024, jahandarChemicalFingerprintsDwarfs2025}. Using a grid of PHOENIX ACES stellar models \citep{Husser_2013} interpolated to $\log g=4.75$ \citepalias[for GJ~3090, $\log g=4.727\pm0.029$;][]{AlmenaraBonfils2022} and convolved to the resolution of NIRPS, we fit a selection of strong spectral lines through $\chi^2$ minimization, yielding $T_\mathrm{eff}=3709\pm34$\,K. The abundances of specific chemical elements are then measured by fitting individual spectral lines for a fixed $T_\mathrm{eff}$ of 3700\,K. We report the average abundances obtained from those fits in Table~\ref{tab:stellar_abundances}, including the overall metallicity of $\mathrm{[M/H]}=0.03\pm0.08$, computed based on the average abundance of all the measured elements \citep{jahandarComprehensiveHighresolutionChemical2024}. We note the relative depletion of OH, which was observed in a previous sample of 31 M dwarfs observed in the near-infrared \citep{jahandarChemicalFingerprintsDwarfs2025}. It is still unknown whether this is due to an unusual depletion of oxygen in M dwarfs compared to FGK stars or a systematic underestimation of OH lines in the models. From the NIRPS spectrum, we measure the abundances of key refractory elements composing the bulk of planetary cores and mantles (Fe, Mg, and Si). The molar iron-to-magnesium ratio is $\mathrm{Fe/Mg=0.69\pm0.16}$, in agreement with the solar value \citep[$0.79\pm0.10$;][]{Asplund2009}.

Alternatively, we derived \teff\ and metallicity (\feh) from the measurements of selected lines in the combined HARPS spectrum by using the machine learning tool {\tt ODUSSEAS}\footnote{\url{https://github.com/AlexandrosAntoniadis/ODUSSEAS}} \citep{Antoniadis20,Antoniadis24}. This code applies a machine learning model trained with the same lines measured and calibrated in a reference sample of 47 M dwarfs observed with HARPS for which their \feh were obtained from photometric calibrations \citep{Neves12} and their \teff\ from interferometric calibrations \citep{Khata21}. With this method, we derived a \teff\,=\,3691\,$\pm$\,95\,K and \feh=\,0.02\,$\pm$\,0.11\,dex, in excellent agreement with the values found with NIRPS spectrum.

The adopted \teff~and \feh are taken as the weighted average of the NIRPS and HARPS measurements, yielding $T_\mathrm{eff}=3707\pm32$\,K and $\mathrm{[Fe/H]}=0.12\pm0.06$\,dex. When compared to the measurements of \citetalias{AlmenaraBonfils2022}, we find an agreement with the \feh value ($-0.06\pm0.12$\,dex) while the effective temperature has a 2$\sigma$ discrepancy ($3556\pm70$\,K). Still, our measurements are both consistent with recent values obtained with \texttt{ODUSSEAS} using spectra from FEROS \citep[$T_\mathrm{eff}=3737\pm104$\,K and $\mathrm{[Fe/H]}=0.04\pm0.13$\,dex;][]{Antoniadis24}.

\begin{table}[h]
    \centering
    \caption{GJ~3090 stellar abundances measured with NIRPS}
    \begin{tabular}{lcc}
        \hline
        \textbf{Element} & \textbf{[X/H]}$^*$ & \textbf{\# of lines} \\ 
        \hline
        Fe I &  0.16 $\pm$ 0.07 & 10 \\
        Mg I &  0.22 $\pm$ 0.05 & 5  \\
        Si I &  0.00 $\pm$ 0.17 & 1  \\
        Ca I & -0.11 $\pm$ 0.39 & 3  \\
        Ti I &  0.14 $\pm$ 0.11 & 14 \\
        Al I &  0.20 $\pm$ 0.10 & 2  \\
        C I  &  0.30 $\pm$ 0.17 & 1  \\
        K I  & -0.10 $\pm$ 0.17 & 1  \\
        OH   & -0.51 $\pm$ 0.03 & 39 \\ 
        \hline
        [M/H] & 0.03 $\pm$ 0.08 & --- \\
        \hline
        \multicolumn{3}{l}{$^*$Molar relative-to-solar abundances}
    \end{tabular}
    \label{tab:stellar_abundances}
\end{table}

For the stellar mass ($M_*$) and radius ($R_*$) of GJ~3090, we adopt the values of \citetalias{AlmenaraBonfils2022} which are computed from the empirical relations of \citet{Mann2015,Mann2019},  calibrated on nearby M dwarfs with interferometric radii and dynamical masses, and based on the absolute $K_s$-band magnitude from 2MASS \citep{Skrutskie2006}.

We find that the effective temperature of GJ~3090 derived in this work ($T_{\mathrm{eff}} = 3707 \pm 32\,\mathrm{K}$) is in better agreement with the empirical radius and mass ($0.516 \pm 0.016 \, R_*$ and $0.519 \pm 0.013 \, M_*$) than the temperature from \citetalias{AlmenaraBonfils2022} ($T_{\mathrm{eff}} = 3556 \pm 70\,\mathrm{K}$). Using the Stefan–Boltzmann law with the luminosity from the SED fitting of \citetalias{AlmenaraBonfils2022} ($L_* = 0.0455^{+0.0018}_{-0.0016} \, L_{\odot}$), the radius predicted from our $T_{\mathrm{eff}}$ differs from the \citet{Mann2015} empirical radius by only $0.006\sigma$, whereas the radius predicted from the \citetalias{AlmenaraBonfils2022} temperature is off by $1.89\sigma$. Converting these radii into masses using the radius–mass relations of \citet{Boyajian2012}, our temperature yields a mass prediction that differs from the \citet{Mann2019} value by $1.59\sigma$, compared to $2.76\sigma$ when using the \citetalias{AlmenaraBonfils2022} temperature.

\section{Photometry analysis}\label{sect:photometry_analysis}
\subsection{Transit analysis with TESS}\label{sect:tess_analysis}
We performed the modelling of the TESS photometric data using the \texttt{juliet} framework \citep{Espinoza2019_juliet}, which combines several open-source tools for joint analysis of transits, radial velocities, and GPs. Transit modelling is carried out with \texttt{batman} \citep{Kreidberg2015_batman}. For Gaussian process regression, it employed either \texttt{george} \citep{Ambikasaran2015_george} or \texttt{celerite} \citep{Foreman-Mackey2017_celerite}. \texttt{juliet} is using the nested sampling algorithm implemented in \texttt{dynesty} \citep{Speagle2020_dynesty}, it explores the parameter space by sampling from the priors rather than optimizing a likelihood function.

The transit model simultaneously fits for the stellar density $\rho_\star$, planetary parameters, and photometric jitter. Instead of fitting directly for the planet-to-star radius ratio ($p = R_p / R_\star$) and the orbital impact parameter ($b = a/R_\star \cos i$), we followed the parametrization proposed by \citet{Espinoza2018}, which involves fitting for the variables $r_1$ and $r_2$. This ensures a more complete and physically consistent exploration of the ($p$, $b$) parameter space. The explicit analytical expressions relating $(r_1, r_2)$ to $(p, b)$ are given in detail in \citet{Espinoza2018} as well as in the documentation of the \texttt{juliet} package.

We used the power-2 limb-darkening law implemented within \texttt{juliet} by \citet{Parc2025}, which has been shown to best reproduce intensity profiles of cool stars \citep{Morello2017}. The corresponding limb-darkening coefficients and their uncertainties for each photometric filter were computed using the \texttt{LDCU} tool\footnote{\url{https://github.com/delinea/LDCU}} \citep{Deline2022}. This tool, a refined version of the routine by \citet{Espinoza2015}, calculates limb-darkening coefficients along with uncertainties by propagating the errors on stellar parameters through grids of stellar intensity profiles derived from ATLAS \citep{Kurucz1979} and PHOENIX \citep{Husser_2013} synthetic spectra. The resulting coefficients were used as Gaussian priors in our transit fit. Since the TESS PDCSAP flux is already corrected for contamination, we fixed the dilution factor to unity. To account for possible underestimation of the photometric uncertainties due to additional systematics, we added in quadrature a jitter term $\sigma_{i}$ to the reported uncertainties. Additionally, to model the residual photometric variability, we included a GP component for each sector, using a Mat\'{E}rn 3/2 kernel. The GP is characterized by two hyperparameters: the amplitude ($\sigma_{GP}$) and the characteristic timescale ($\rho_{GP}$).
We assumed a circular orbit so we fixed the eccentricity
to zero and used normal priors around the ExoFOP values for
the period and transit epoch. We used a normal prior for stellar density using the stellar mass and radii values from \citetalias{AlmenaraBonfils2022}.

\begin{figure}[ht]
    \centering
    \includegraphics[width=\columnwidth]{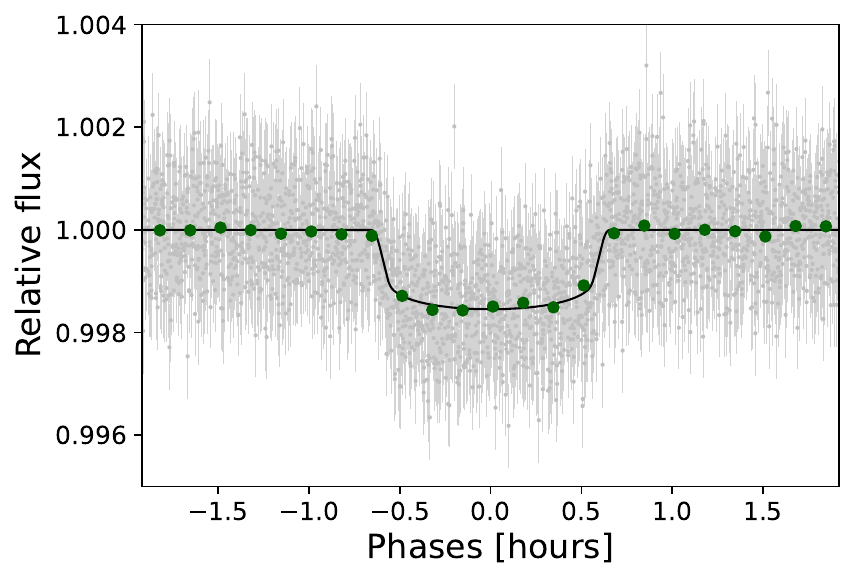}
    \caption{Phase-folded TESS light curves of GJ 3090~b (grey points). The dark green circles are data binned to 10 min. The black lines represent the median model of each sector from the fit.}
    \label{fig:LC_TESS}
\end{figure}

Fig.~\ref{fig:full_sectors_LCs_TESS} shows the four TESS sectors with the resulting model obtained with the \texttt{juliet} fit and Fig.~\ref{fig:LC_TESS} shows the phase-folded light curve of GJ 3090~b. The prior and posterior distributions are shown in Table~\ref{tab:priors_posteriors_tess_fit}.

We derived a radius of 2.18$\pm$0.06~$R_\oplus$ for GJ 3090~b using the stellar radius from \citetalias{AlmenaraBonfils2022}. This is in line with the previous values from \citetalias{AlmenaraBonfils2022} and \citetalias{Ahrer2025}. Moreover, we obtained an inclination of 86.9$\pm$0.2$^\circ$ and an impact parameter of 0.70$\pm$0.05. The period and time of conjunction posteriors of the transit fit is used as priors for the RV fit. 

\subsection{Magnetic cycle with ASAS-SN}\label{cycle discussion}
\begin{figure}[ht]
    \centering
    \hspace{-0.7cm} 
    \includegraphics[width=0.95\linewidth]{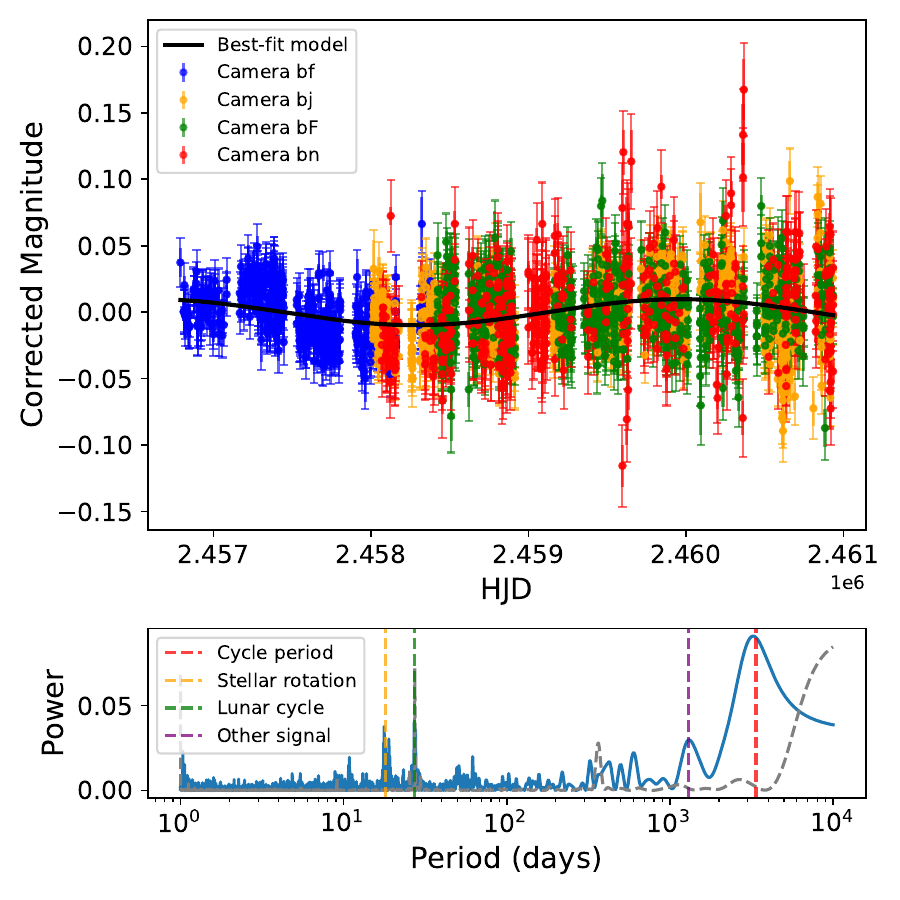}
    \caption{
    \textbf{Top panel:} Joint best-fit sinusoidal model to the ASAS-SN photometric data from all available cameras (\textit{bf}, \textit{bj}, \textit{bF}, and \textit{bn}). The solid lines represent the best-fit model. 
    \textbf{Bottom panel:} Lomb–Scargle periodogram of the ASAS-SN photometry after camera offset corrections. The blue curve shows the power spectrum. The strongest peak corresponds to the best-fit magnetic cycle period of \Pcyc\,days (red dashed line), while the orange dashed line indicates the stellar rotation period of 18.0\,days and the green dashed line shows the sidereal lunar cycle that arises from the applied lunar correction. The grey dashed curve represents the window function.
    }
    \label{fig:ASASN-data}
\end{figure}

The long time span of the HARPS (HAM) dataset ($\sim 6$ years) allows us to scan for long-term signals indicative of long-period planets or magnetic cycles. The first hint of its presence is revealed by the ASAS-SN photometry dataset shown in Figure \ref{fig:ASASN-data}. Also, the second epoch of the archival HARPS data separated by 200 days shows a 6\,m\,s$^{-1}$ offset relative to the first epoch which cannot be explained by a change in the instrument's stability since no major modifications were made on the spectrograph between those epochs. Our multidimensional GP model fails to model this offset which then shows up in the residuals. 

These hints encouraged us to look further into explanations for this long-term signal. To distinguish between the planetary and long-term stellar variability scenarios, we retrieved photometric data from the ASAS-SN survey, as photometry has been shown to exhibit signatures of magnetic activity cycles in stars \citep{SuarezMascareno2016}. We perform Bayesian inference of a sinusoidal model (eq. \ref{cycle model}) on the ASAS-SN dataset which is acceptable for long-term variability because few full cycles are observed. Each available camera from the ASAS-SN network was fitted with independent offset and jitter terms. The coloured points in Figure \ref{fig:ASASN-data} show the magnitudes corrected using the fitted offset of each camera. Contrary to the best-fit model shown in the top panel, the periodogram displayed in the lower panel is computed from the light curve after subtracting only the median value of each camera, ensuring that the periodogram remains model-independent. We used the package \texttt{UltraNest} \citep{Buchner2021} with a reactive nested sampling algorithm, $440$ live points, and a slice sampler performing $44$ steps per iteration, distributed over $20$ CPU cores. The best-fit curve is shown on top of the data in Figure \ref{fig:ASASN-data}. The posterior on the period is \Pcyc\,days. There is another strong signal at $1300$~d in the periodogram seen in Figure \ref{fig:ASASN-data} which could also be a candidate for a long-term trend. In the case of the ASAS-SN fit, the $3370$~d solution is overwhelmingly preferred by the FAP in the periodogram as well as by model comparison ($\Delta \log z \approx 90$, obtained when comparing a model in which the uniform prior upper bound on the period is set to the RV timespan of $\sim$2000 d with another model in which the upper bound is set to the full ASAS-SN timespan of $\sim$4000 d), but we do not exclude the possibility that the lower $1300$~d signal is present in the radial velocities. In order to account for this long-term signal, we decided to add a magnetic cycle model in our main multidimensional fit. We do not fit a magnetic cycle on the TESS photometry since the sectors are too sparse and discontinuous to detect signals on those timescales.

The magnetic cycle amplitudes of all modelled measurements converged to zero, except for the HARPS radial velocities, which yielded a best-fit amplitude of $K_{H}^{\mathrm{cyc}} = 4.14^{+1.07}_{-0.90}$~m/s. This result is not surprising since NIRPS does not yet have the time baseline required to be sensitive to such long-period signals. As for the $\Delta T$ indicator, it does not appear to be sufficiently sensitive or stable enough to capture long-term magnetic trends in our dataset.

As shown by the model comparison presented at the beginning of Section~\ref{sec:results}, a long-term sinusoidal component is statistically favoured. However, given the lack of a significant peak near $\sim$3370~d in the HARPS RV periodogram (see Figure \ref{fig:periodograms}) and the absence of a corresponding signal in the $\Delta T$ indicator, we refrain from assigning a definitive physical interpretation. We therefore regard this variability as a candidate magnetic cycle or, more conservatively, an unclassified long-term stellar trend. It is nonetheless important to model it explicitly; otherwise, the GP would be forced to absorb the long-term variability, compromising the physical interpretability of its hyperparameters and biasing the stellar-activity component of the fit \citep{Stock2023}.

\section{Radial velocity modelling and inference} \label{sec:modelling}
We denote all measurements by $y_{s,q}$, their associated uncertainties by $\Delta y_{s,q}$, and their times of measurement by $t_{s,q}$, where the subscript $s$ refers to the data source (either NIRPS, HARPS HAM, HARPS EGGS or TESS) and the subscript $q$ refers to the type of quantity measured (either visible RVs, infrared RVs, $\Delta T$ or Flux).

The measurements are modelled as the sum of a systematic offset ($\gamma$), a stellar variability model, which include both rotational modulation ($Rot$) and a potential long-term magnetic cycle ($Cyc$), and a Keplerian model ($Planet$) which is only applied on radial velocities. The rotational modulation is modelled using a multidimensional GP on all physical quantities ($q$). The GP is trained simultaneously on the TESS photometry, the $\Delta T$ activity indicator obtained from the HARPS HAM spectra, and the radial velocities of NIRPS (infrared) and HARPS (visible) which sum up to a 4D GP. In addition to accounting for rotationally induced signals, we model a possible long-term cycle by including a single periodic component. To test whether this additional long-term component is statistically justified, we compared two models: one including all components except the long-term sinusoid ($\gamma \, + Rot + Planet$) and one including all components including the sinusoid ($\gamma \, + Rot + Cyc + Planet$). The model with the long-term component is favoured with $\Delta \ln Z = 10.3$. Equation \ref{model components} shows the model components for all physical quantities $q$ and the next sections go through how each of the components are modelled:

\begin{equation}
\begin{aligned}
\Delta F_{\text{TESS}} &= \gamma \, + Rot,\\
\Delta T_{\text{HARPS}} &= \gamma \, + Rot + Cyc, \\
\Delta RV_{\text{NIRPS}} &= \gamma \, + Rot + Cyc + Planet, \\
\Delta RV_{\text{HARPS}} &= \gamma \, + Rot + Cyc + Planet. \\
\end{aligned} \label{model components}
\end{equation}

\subsection{Source offsets and instrument jitters}
Zero-point offsets ($\gamma$) and jitters ($\sigma_{\text{jit}}$) are included on all $y_{s,q}$ and $\Delta y_{s,q}$ respectively:

\begin{align}
    y_{s,q} \rightarrow& \, y_{s,q} - \gamma_{s,q} \\ 
    \Delta y_{s,q} \rightarrow& \, \sqrt{\Delta y_{s,q}^2 + \sigma_{\text{jit};s,q}^2} \quad \forall \, s.
\end{align}

Each data source ($s$) and observable ($q$) is assigned its own offset parameter and jitter term to align the datasets consistently and model excess noise or potential instrumental drifts not captured by the model, preventing underestimated uncertainties.

\subsection{Magnetic cycle correction}
The long-term magnetic cycle of period $P_{cyc}$ is modelled by including a single periodic component, with each time series ($q$) having its own cycle epoch ($\tau^{cyc}$) and amplitude ($K^{cyc}$) parameters to account for differences in how the cycle manifests across observables. 

\begin{align} \label{cycle model}
        y_{s,q} \rightarrow y_{s,q} - K^{cyc}_q \sin \left(2 \pi \, \frac{t_{s,q} - \tau^{cyc}_{q}}{P_{cyc}}\right) \quad \forall \, q.
\end{align}

This correction is crucial for disentangling planetary signals from long-term stellar activity trends. $P_{cyc}$ is informed by the continuous ASAS-SN photometry data which shows a long-term signal. 

\subsection{Planetary signals}

The radial velocity variations induced by planetary companions are modelled using the Keplerian equation, which describes the reflex motion of the host star under the gravitational influence of orbiting planets. This component of the model applies only to time series $q$ belonging to the radial velocity subset, denoted $\mathbf{RV} = \{\Delta \mathrm{RV}_{\text{HARPS}}, \Delta \mathrm{RV}_{\text{NIRPS}}\}$. The full Keplerian model is given by:

\begin{align}\label{eq:keplerian_model}
y_{s,q} \rightarrow y_{s,q} - K \left[\cos(\omega + \nu(t)) + e \cos(\omega)\right] \quad \forall q \in \mathbf{RV},
\end{align}

where $K$ is the RV semi-amplitude, $\omega$ is the argument of periastron, $e$ is the orbital eccentricity, and $\nu(t)$ is the true anomaly at time $t$. This formulation captures the full Keplerian motion, allowing for eccentric orbits.

In cases where eccentricity is negligible or unconstrained by the data, a circular orbit approximation can be adopted. The model then simplifies to:
\begin{align}
y_{s,q} \rightarrow y_{s,q} + K \sin \left(2 \pi \frac{t_{s,q} - t_0}{P} \right) \quad \forall q \in \mathbf{RV},
\end{align}
where $t_0$ is the time of inferior conjunction and $P$ is the orbital period. This approximation requires a change in sign relative to eq. \ref{eq:keplerian_model} (\citet{Stefanov2025}, hereafter \citetalias{Stefanov2025}). We note that the Keplerian model does not take planet-planet interactions into account.

\subsection{Stellar-activity modelling}
Gaussian processes are the basis of the stellar activity modelling in this work. GPs are non-parametric models that provide a flexible framework for capturing complex variability without assuming a specific functional form \citep{Rasmussen2005}. Formally, a GP defines a distribution over functions such that any finite set of points drawn from it follows a joint multidimensional Gaussian distribution. GPs are particularly effective for modelling stellar variability in time-series data, as they do not require prior knowledge of the distribution or evolution of active regions on the stellar surface. However, their flexibility also poses a risk: without careful modelling, GPs can overfit the data and inadvertently absorb astrophysical signals such as planetary signatures \citep{Damasso_2019, Blunt_2023}. Stellar activity can generally be described as a quasi-periodic signal.

To model and correct the correlated noise introduced in the RV time series by stellar rotation, we use the multidimensional Gaussian process framework \citep{Rajpaul2015}. This state-of-the-art technique allows for the simultaneous fit of multiple time series using an underlying GP and its derivative following a scheme similar to the $FF'$ formalism \citep{Aigrain2012}. In order to model this multidimensional GP, we use the \texttt{S+LEAF} package \citep[hereafter \citetalias{Delisle2022}]{Delisle2022}, which extends the semi-separable matrix formalism introduced by \texttt{celerite} \citep{celerite1} to enable rapid evaluation of GP models, even for large datasets. This computational efficiency is crucial for handling extensive data while maintaining flexibility.

To maintain the linear scaling of the likelihood computation made possible by the semi-separable framework, \texttt{S+LEAF} provides a set of computationally efficient kernels that approximate the behaviour of more general GP kernels, such as the commonly used quasi-periodic kernel. These approximations preserve the key physical characteristics of stellar activity signals while allowing for fast evaluation. In this study, we tested three such kernels available in \texttt{S+LEAF} \citepalias{Delisle2022}: Double SHO, Mat\'{E}rn 3/2 exponential periodic (MEP), and exponential-sine periodic (ESP). These kernels differ in their mathematical complexity and approach to modelling quasi-periodic stellar activity. The Double SHO represents the simplest method, using two simple harmonic oscillators with basic trigonometric functions to capture the stellar rotation period and its first harmonic. The MEP kernel increases complexity by combining a Mat\'{E}rn-3/2 kernel with harmonic oscillators, employing exponential decay functions to better represent the quasi-periodic nature of stellar signals. The ESP kernel is the most sophisticated, utilizing modified Bessel functions to weight multiple harmonics, enabling it to model intricate periodic structures and complex quasi-periodic behaviour.

Equation \ref{FF'} defines how the underlying GP is modelled on each physical quantity $q$:

\begin{equation}
    \begin{aligned}
    \Delta RV_{\text{HARPS}} &= \alpha_0\, k + \beta_0\, k', \\
    \Delta RV_{\text{NIRPS}} &= \alpha_1\, k + \beta_1\, k', \\
    \Delta T_{\text{HARPS}} &= \alpha_2\, k, \\
    \Delta F_{\text{TESS}} &= \alpha_3\, k, \\
    \end{aligned} \label{FF'}
\end{equation}

\noindent where $k$ is the underlying GP and $k'$ is its derivative. The free parameters $\alpha$ and $\beta$ scale the contribution of the GP and its derivative to each observable. The first order of the GP ($k$) represents activity-induced variations in photometry-like indicators, specifically the TESS light curve and the $\Delta T$ temperature indicator, while the stellar contribution to the radial velocities is modelled using both the first ($k$) and second order ($k'$) of the same GP. The first-order term generally corresponds to brightness variations caused by the contrast between star spots and the stellar surface as well as suppression of convective blueshift in magnetized regions (spots), while the second-order term accounts for the Doppler shifts induced by these spots as they move across the stellar disk due to rotation \citep{Rajpaul2015,Aigrain2012}.

A key modelling choice in our analysis is to restrict the $\Delta T$ indicator to only the first-order GP component. This decision is motivated by both physical and empirical considerations. Physically, as in photometry, the temperature primarily traces the luminous contrast on the stellar surface rather than the Doppler shifts associated with moving regions, making it more appropriate to model $\Delta T$ with a first-order term alone. When allowing both the zeroth-order amplitude ($\alpha_2$) and the derivative amplitude ($\beta_2$) to vary, the posteriors exhibited a symmetry between positive and negative values for these amplitudes, leading to a degeneracy in the solution without improving the fit quality. Restricting the model to only the zeroth-order amplitude removed this degeneracy and still provided an excellent fit to the data, further justifying its exclusion from the final model.

A key advantage of the multidimensional approach is that it mitigates the risk of misattributing planetary power to stellar activity compared to a serial GP approach, where a GP is first trained on activity indicators and then its posterior is used as a prior for the RV fit \citep[Section 4.1.3;][]{Tran2023}. In the serial approach, the GP is optimized separately for each dataset, which can lead to an overestimation of the activity component in the RVs, as the GP has the flexibility to fit any residual variations, including planetary signals or magnetic cycles. This can result in the suppression or distortion of true planetary signals. Conversely, in the multidimensional approach, all datasets (RVs and activity indicators) are modelled jointly within the same covariance matrix, forcing the GP component to explain correlated variations consistently across all observables. This approach has been successfully used to confirm challenging Keplerian signals around highly active stars \citep{Jones2022, Barrag_n_2023, GonzlezHernndez2024}.

\subsection{Choice of priors}
All priors can be found in Table \ref{tab:priors_posteriors}. For the planetary orbit parameters $e$ and $\omega$, we used uniform priors on $\sqrt{e} \cos \omega$ and $\sqrt{e} \sin \omega$ between -1 and 1, as this has been shown to better explore the parameter space \citep{Lucy1971}. As the period of GJ~3090\,b is more precisely constrained by the transit data than by the radial-velocity data, we chose to use the posteriors from the transit analysis (see Section~\ref{sect:tess_analysis}) for the period and time of transit as priors in our analysis, since the additional TESS sector available since the system’s discovery has improved the constraints on these parameters. To search for additional planets, we initially ran a model with broad priors on the period ($\mathcal{U}(3, 200)$) to conduct a blind search. We then used the posterior from this fit to define a more constrained uniform prior for a second fit, allowing for improved posterior sampling and more robust Bayesian evidence measurements \citep{Aitken2013}. We use a normal prior on $P_{cyc}$ centered on the 3370~d periodicity identified by ASAS-SN (see Section \ref{cycle discussion}) with an inflated uncertainty to include the $1300$~d solution at $3 \, \sigma$. The rest of the parameters were given large uniform priors except for jitter parameters and GP or cycle amplitudes which were given log-uniform priors as prescribed in \cite{Stock2023}. 

\subsection{Parameter inference}
We optimized the model parameters using Bayesian inference via nested sampling \citep{Skilling2004,Skilling2006} with the \texttt{dynesty} package \citep{Speagle2020}. To ensure thorough exploration of the parameter space, we used a high number of live points, setting \( \texttt{nlive} = 200 \times N_{dim} \), where \( N_{dim} \) is the number of free parameters in the model. This high-density sampling improves the accuracy of posterior estimations, which are crucial for model comparison.

We employed the multi-ellipsoid bounding method (\texttt{bound=`multi'}) to efficiently constrain the parameter space and reduce computational overhead. Sampling was performed using the random walk algorithm (\texttt{sample=`rwalk'}), which is well-suited for handling correlated parameter distributions, with an increased number of MCMC walk steps ($\text{walks}=10 \times N_{dim}$) to enhance convergence.

\section{Results and discussion} \label{sec:results}

\subsection{Determining the number of planets in the GJ~3090 system}\label{sec:model comparison}

In addition to the known transiting planet GJ~3090\,b, several significant signals appear in the periodogram of the combined RV dataset (top panel of Figure~\ref{fig:periodograms}). The most prominent peak occurs at 16 days, close to the estimated 18-day stellar rotation period \citepalias{AlmenaraBonfils2022}. Another noticeable signal appears near 13 days, which was first reported by \citetalias{AlmenaraBonfils2022} as a potential planetary candidate.

\begin{figure}[h]
    \centering
    \includegraphics[width=\columnwidth]{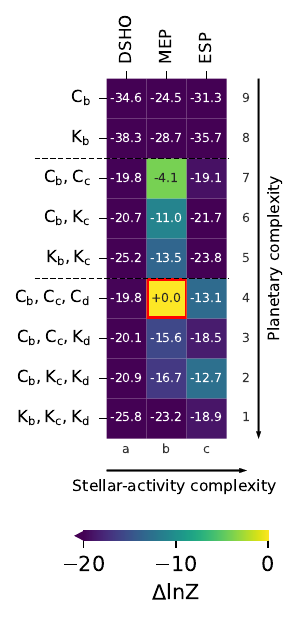}
    \caption{
        Heat map similar to that in \citetalias{Stefanov2025} showing the comparison of Bayesian evidence ($\Delta \ln Z$) for different models. 
        The x-axis displays different stellar-activity kernels in increasing order of complexity, while the y-axis shows different planetary configurations. Circular orbits ($e=0$) are labelled `C', while Keplerian orbits ($e \geq 0$) are labelled `K'. 
        The red rectangle highlights the best model based on the maximum Bayesian evidence. 
    }
    \label{fig:bayesian_evidence_heatmap}
\end{figure}

To assess the most plausible planetary configuration for the system, we conducted a rigorous Bayesian model comparison using the nested sampling log-evidence ($\ln Z$). Figure~\ref{fig:bayesian_evidence_heatmap} displays the $\Delta \ln Z$ values for all tested combinations of stellar activity models (GP kernels) and planetary configurations similarly to \citetalias{Stefanov2025}. Each row corresponds to a planetary model with an increasing number of Keplerian signals, while each column represents a different GP kernel. We adopt the notation $C$ for circular orbits ($e=0$) and $K$ for Keplerian models with free eccentricity ($e \geq 0$). The colour scale encodes the log-evidence relative to the best overall model, offering a joint view of stellar and planetary model complexity.

\begin{figure*}[h]
    \centering
    \includegraphics[width=\textwidth]{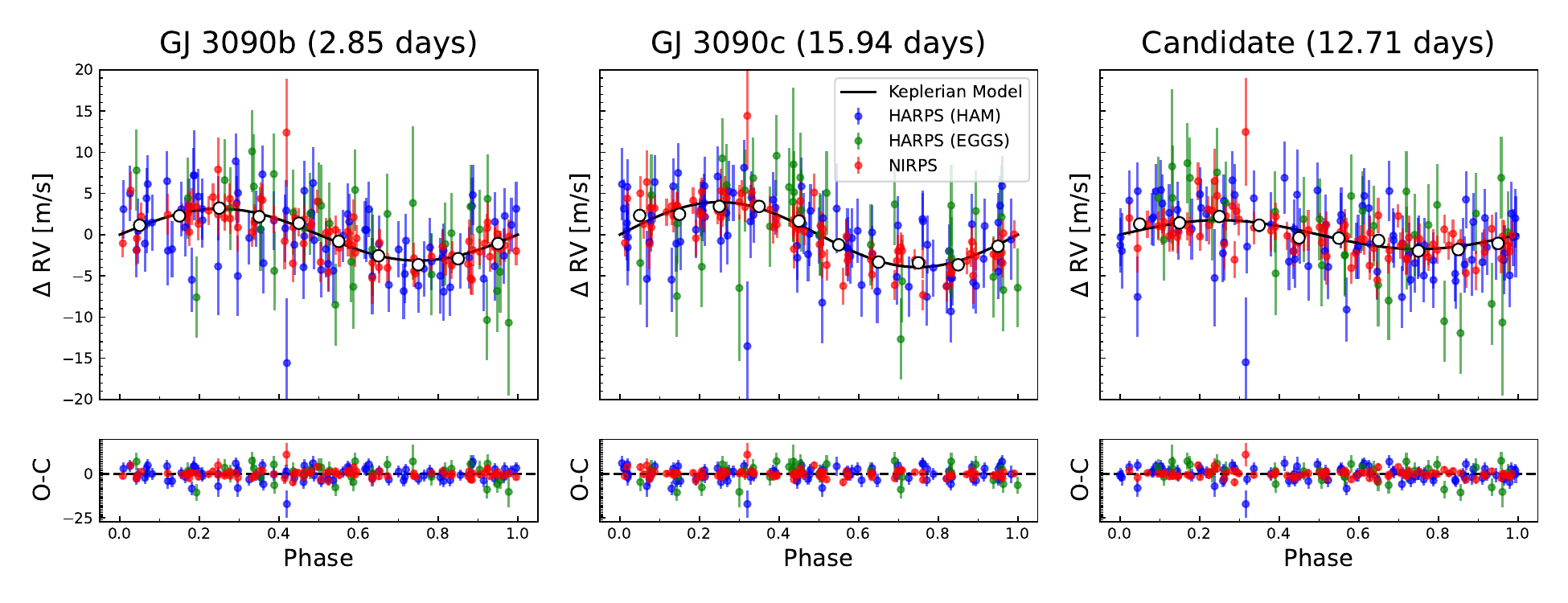}

    \caption{
        Phase-folded radial velocity measurements for the confirmed and candidate planets in the GJ~3090 system. 
        The first panel shows the confirmed planet GJ~3090\,b (2.85 days), the second panel presents the confirmed sub-Neptune GJ~3090\,c (15.94 days), and the third panel displays the 12.7-day candidate planet. 
        The data points from HARPS HAM, HARPS EGGS, and NIRPS are shown with their associated uncertainties, alongside the Keplerian model curves (black lines). 
    }
    \label{fig:phase_folds}
\end{figure*}

Our highest-evidence model (labelled b4) combines three circular Keplerian signals ($C_b$, $C_c$, and $C_d$) with periods of 2.85 days, 15.94 days, and 12.71 days respectively, plus a \bestkernel kernel for stellar activity. The phase-folded radial velocity curves for all three planets are shown in Figure~\ref{fig:phase_folds}, illustrating their individual contributions. Figure \ref{fig:timeseries} displays the GP model prediction overlaid on both the RV datasets and activity indicators. The top-right panel shows the full model (Keplerian plus GP) on the most recent epoch of observations, demonstrating excellent visual agreement with the data in particular for NIRPS. The median residual values after subtracting the full model are $1.15$~m\,s$^{-1}$ for NIRPS, $2.68$~m\,s$^{-1}$ for HARPS (HAM), and $3.65$~m\,s$^{-1}$ for HARPS (EGGS).

To quantify how the model evidence evolves with planetary multiplicity, we compute $\Delta \ln Z$ in three different ways: Let $\{\ln Z_k\}_{i}$ denote the set of log-evidence for planetary model $i$ across all tested kernels $k$. First, we compute the average log-evidence difference (see eq. \ref{eq:dlnz_avg}) between two configurations $i$ and $j$ where $N_k$ is the total number of tested GP kernels. This provides a global view of how favourable one planetary model is over another across all GP kernels. Second, we adopt a more conservative approach by computing the pessimistic (pes) log-evidence (see eq. \ref{eq:dlnz_pes} which compares the worst-performing kernel of the better Keplerian model $i$ with the best-performing kernel of the worse Keplerian model $j$. This illustrates the impact of the choice of kernel on planetary detection. Finally, we compute the best log-evidence difference (see Eq. \ref{eq:dlnz_best}), which compares the log-evidence obtained with the overall best-performing kernel (\bestkernel in our case) for both planetary architecture. This metric represents the most realistic assessment of the models’ relative performance, as it isolates the comparison to the kernel that best captures the stellar activity component. In practice, this best log-evidence is typically the one reported when confirming an exoplanet candidate, since once a kernel is shown to perform significantly better, the planetary configuration should be evaluated on that basis alone.

\begin{align}
    (\Delta \ln Z)_{\text{avg}} &= \frac{1}{N_k} \sum_{k=1}^{N_k} \left( \ln Z_{k, i} \right)
    - \frac{1}{N_k} \sum_{k=1}^{N_k} \left( \ln Z_{k, j} \right), \label{eq:dlnz_avg} \\[6pt]
    (\Delta \ln Z)_{\text{pes}} &= \min_{k} \left( \ln Z_{k, i} \right)
    - \max_{k} \left( \ln Z_{k, j} \right), \label{eq:dlnz_pes} \\[6pt]
    (\Delta \ln Z)_{\text{best}} &= \best_{k} \left( \ln Z_{k, i} \right)
    - \best_{k} \left( \ln Z_{k, j} \right). \label{eq:dlnz_best}
\end{align}

All $\Delta \ln Z$ are then converted into equivalent sigma significances using Equations 8 and 9 of \citet{Benneke2013}.

The two-planet model, including GJ~3090\,c at 16 days, is decisively favoured over the single-planet model, with $\sigma_{\text{avg}} = 6.0$, $\sigma_{\text{pes}} = 3.5$, and $\sigma_{\text{best}} = 6.7$. The second planet is detected with high statistical confidence ($>5\sigma$), justifying its classification as a confirmed planet. Its proximity to the stellar rotation period motivates a dedicated discussion in Section~\ref{sec:proximity to stellar rotation} to exclude stellar activity as its origin.

The three-planet model, which incorporates the 12.7-day signal as a candidate, receives modest additional support compared to the two-planet model with $\sigma_{\text{avg}} = 3.1$ and $\sigma_{\text{best}} = 3.3$ while the pessimistic metric prefers the two-planet model by $\sigma_{\text{pes}} = 6.0$. If we add the 12.7-day planet as a second planet and compare it with the models containing only GJ~3090\,b, we get $\sigma_{\text{best}} = 4.4$. This warrants caution and indicates that additional data or improved stellar activity modelling may be necessary to establish the planetary nature of this signal.

We tested a four-planet model by adding a keplerian signal with an open prior on the period ($\mathcal{U}(3, 200)$) even though there were no significant peaks in the periodogram of the residuals. We got no clear convergence to a planetary signal. 

To quantify the improvement brought by NIRPS, we compare the semi-amplitude significance $N_\sigma = K / \sigma_K$ of the planetary signals in the HARPS-only and combined NIRPS+HARPS fits (see Table~\ref{tab:planet_posteriors}). For GJ~3090~b, $N_\sigma$ increases from $\sim5.5$ (HARPS-only) to $\sim9.8$ in the combined fit, and for GJ~3090~c from $\sim4.7$ to $\sim7.7$, demonstrating a clear gain in precision. For the 12.7-day candidate, however, there is no gain: the HARPS-only and NIRPS-only semi-amplitudes differ significantly, and their combination does not improve the detection ($N_\sigma \sim 4.0$).

Finally, as seen in Figure \ref{fig:bayesian_evidence_heatmap}, eccentric models are consistently disfavoured by Bayesian evidence, suggesting that the eccentricities of the planets are likely low and not sufficiently constrained by the available data. As a result, we adopt model b4—which includes three circular Keplerian signals and employs the MEP kernel—as the baseline model for the remainder of the discussion.

\subsection{Explaining the absence of outer planet transits}\label{sec:additionnal transits}

Assuming coplanar orbits aligned with the inclination of GJ~3090\,b (\inclination \, deg) updated in this work, we computed the expected impact parameters of the outer planets using their semi-major axes and the stellar radius. The resulting values are $b_c = 2.22 \pm 0.14$ for GJ~3090\,c and $b_{\text{cand1}} = 1.91 \pm 0.12$ for the 12.7-day candidate. Since both values are significantly greater than 1, the planetary orbits do not intersect the stellar disk from our line of sight. This geometrical configuration naturally explains why only GJ~3090\,b is observed in transit, while the outer planets remain undetected in photometry. The non-detection of GJ~3090\,c and the 12.7-day candidate is shown in Figure~\ref{fig:phase_folded_light_curves}, which plots the phase-folded TESS light curves folded on the best-fit ephemeris of the two outer planets.

To further assess the detectability of potential transits, we used the \texttt{spright} package \citep{Parviainen2023} to estimate the expected transit depths based on probabilistic mass–radius relations inferred from the distribution of known exoplanets. For GJ~3090\,c, the predicted depth is $2774^{+1135}_{-804}$~ppm, well above the TESS detection threshold. For the 12.7-day candidate (GJ~3090\,d), the estimated depth depends on its nature: if it is a super-Earth, the expected depth is $679^{+97}_{-89}$~ppm, whereas if it is a sub-Neptune, the depth increases to $1171^{+459}_{-378}$~ppm. Since its inferred mass is close to those of exoplanets lying in the radius valley, both compositional scenarios are plausible. However, in all cases, such depths would have been detectable in the available TESS photometry. This supports the interpretation that the absence of transit signals is due to geometric misalignment rather than insufficient signal strength.

\subsection{New constraints on GJ~3090\,b}\label{sec:GJ3090b new constraint}

Using the combined HARPS and NIRPS radial velocity datasets which include 152 new RV measurements since \citetalias{AlmenaraBonfils2022}, we present updated constraints on the mass, radius and bulk density of GJ~3090\,b. To get an absolute value for the mass, we use the inclination measured. Our analysis yields a mass of \massb \, $M_{\oplus}$, while \citetalias{AlmenaraBonfils2022} reported a value of $3.34 \pm 0.72,M_{\oplus}$. The new mass is in good agreement with a reanalysis of the archival HARPS dataset made by \citet{Osborne2024}, hereafter \citetalias{Osborne2024}, that found $4.59_{-1.04}^{+1.13} M_{\oplus}$ by fitting a multidimensional GP model on RVs, FWHM (Full Width Half Maximum) and BIS (Bisector Inverse Slope). The mass obtained in this work is derived from the posterior of our highest-evidence model (b4), which includes three circular planets and uses the \bestkernel kernel for activity modelling. The relative error on the mass of GJ~3090\,b thus went from $22$\% ($4.6 \sigma$, \citetalias{AlmenaraBonfils2022}), reevaluated at $24$\% ($4.2 \sigma$, \citetalias{Osborne2024}) and finally to $10$\% ($9.6 \sigma$, this work). The high precision on the mass of GJ~3090\,b achieved in this work is primarily attributable to the GP modelling, the extended temporal baseline of the combined datasets, and the enhanced precision of the NIRPS measurements. 

\begin{figure}[h]
    \centering
    \includegraphics[width=\columnwidth]{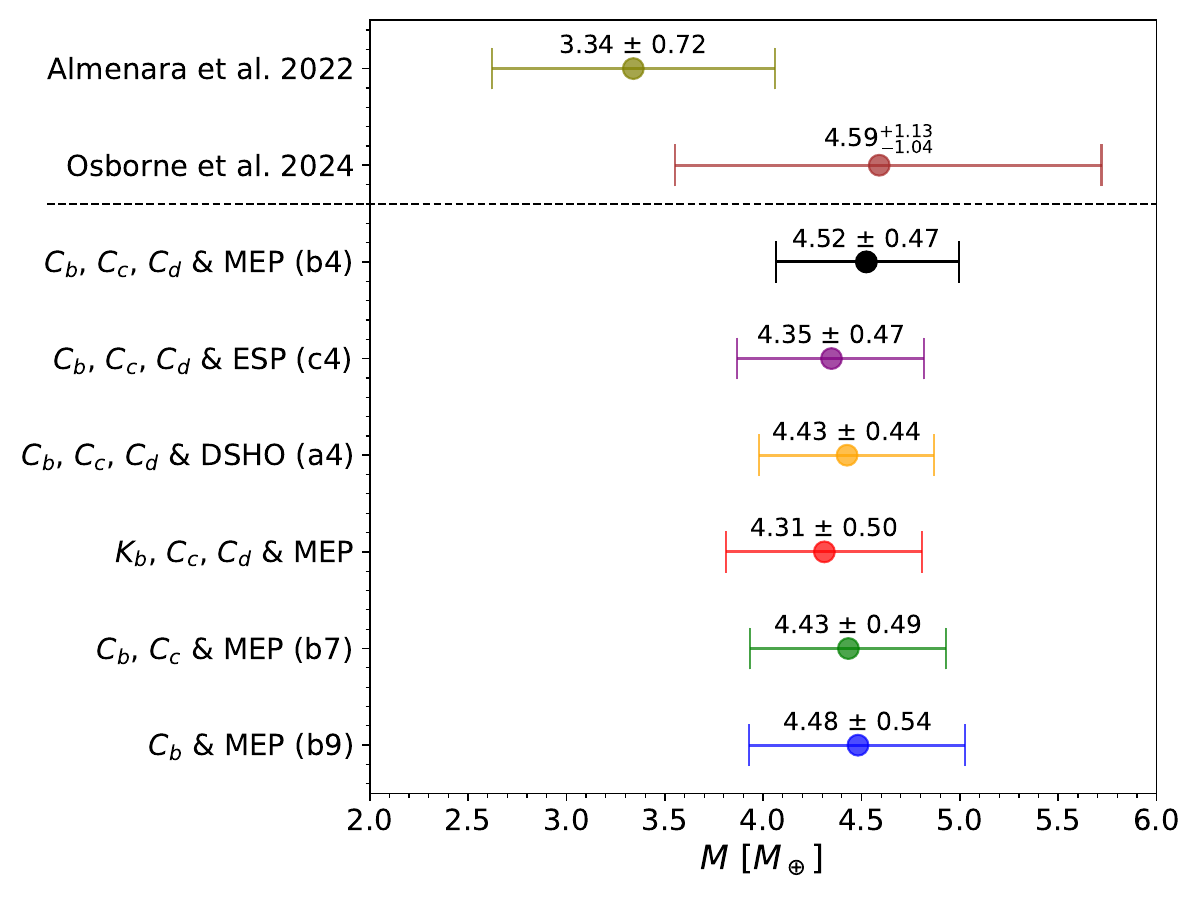}
    \caption{Mass estimates of GJ~3090\,b under different planetary configurations and different stellar-activity kernels. The models range from single-planet solutions (GJ~3090\,b only) to multi-planet scenarios, including the 16-day confirmed planet (GJ 3090\,c) and candidate planet at 13 days. The three-planet models are also shown with different kernels and a model with a non-zero eccentricity on GJ~3090\,b. The adopted model is shown in black. The mass constraints from \citetalias{AlmenaraBonfils2022}, \citetalias{Osborne2024}, and an independent reanalysis of the 55 archival HARPS data points are also shown. A dashed horizontal line separates the models tested in this work from previously published mass estimates and a re-analysis of the HARPS archival data.}
    \label{fig:mass_comparison}
\end{figure}

To ensure that the inferred mass of GJ~3090\,b is robust against modelling choices, we compare its posterior distribution across five different model configurations. Three models vary the number of Keplerian signals (1, 2, or 3 planets) while holding the kernel fixed (MEP), and two others vary the kernel type while keeping the three-planet configuration. As shown in Figure~\ref{fig:mass_comparison}, the resulting mass posteriors are consistent well within $1\sigma$, indicating that neither the inclusion of additional planets nor the choice of GP kernel significantly affects the mass constraint on GJ~3090\,b.

\citetalias{AlmenaraBonfils2022} previously measured a radius of $2.13 \pm 0.11\,R_{\oplus}$ from 3 TESS sectors, which, combined with their mass estimate, implied a bulk density of $1.89^{+0.52}_{-0.45}$ g\,cm$^{-3}$. More recently, a JWST transmission spectroscopy study of the planet \citepalias{Ahrer2025} refined the radius measurement to $2.171 \pm 0.068\,R_{\oplus}$ which is compatible with our re-analysis of the 4 available TESS sectors which yielded a radius of \radius \, $R_{\oplus}$. When combined with our updated mass, this yields a new bulk density of \density g\,cm$^{-3}$.

This revised density places GJ~3090,b closer to the bulk population of known sub-Neptunes orbiting M dwarfs, which typically have mean densities of $\sim$2–3 gcm$^{-3}$ \citep[e.g.][]{Parc2024}. Furthermore, the updated mass and radius yield a lower transmission spectroscopy metric (TSM; \citealt{Kempton_2018}). While \citetalias{AlmenaraBonfils2022} reported a TSM of $221^{+66}_{-46}$, we obtain a revised value of $182^{+27}_{-23}$ using the new parameters derived in this work. This decrease in TSM may partially account for the lack of detectable atmospheric features reported by 
\citetalias{Ahrer2025} and \citet{Parker2025}.

\begin{table*}[ht!]
\small
\setlength{\textwidth}{1pt}
\renewcommand{\arraystretch}{1.2}
\centering
\begin{threeparttable}
\caption{Main planetary parameters of the GJ~3090 system}
\label{tab:planet_parameters}
\begin{tabular}{lp{3.0cm}cccccc}
\toprule
& \textbf{Parameter} 
& \multicolumn{2}{c}{\textbf{GJ~3090\,b}} 
& \multicolumn{2}{c}{\textbf{12.7\,d Candidate}}
& \textbf{GJ~3090\,c} \\
\cmidrule(lr){3-4} \cmidrule(lr){5-6}
& & \citetalias{AlmenaraBonfils2022} & This work & \citetalias{AlmenaraBonfils2022} & This work & \\
\midrule
& Period, $P$ (d) & $2.85310195 \pm 8.0 \times 10^{-7}$ & \period & $12.729^{+0.025}_{-0.031}$ & $12.714^{+0.010}_{-0.076}$ & $15.9407 \pm 0.0058$ \\
& $t_c$ (BJD$-2457000$) & $1370.41849 \pm 0.00034$ & \tc & $1370.96^{+1.2}_{-0.9}$ & $1360.3_{-1.3}^{+1.2}$ & $1350.83 \pm 0.61$ \\
& Semi-major axis, $a$ (au) & $0.03165 \pm 0.00027$ & $0.0316 \pm 0.0003$ & $0.08575 \pm 0.00074$ & $0.0857 \pm 0.0007$ & $0.0997 \pm 0.0008$ \\
& Impact parameter, $b$\tnote{b} & $0.631^{+0.093}_{-0.26}$ & $0.70 \pm 0.05$ & $19^{+10}_{-13}$ & $1.91 \pm 0.12$ & $2.22 \pm 0.14$ \\
& Mass, $M_p$ ($M_{\oplus}$)\tnote{a} & $3.34 \pm 0.72$ & \massb  & $17.1^{+8.9}_{-3.2}$ & \masscandidateone & \massc \\
& Radius, $R_p$ ($R_{\oplus}$) & $2.13 \pm 0.11$ & \radius & -- & -- & -- \\
& Density, $\rho$ (g\,cm$^{-3}$) & $1.89^{+0.52}_{-0.45}$ & \density & -- & -- & -- \\
& Insolation, $S$ ($S_{\oplus}$) & -- & $45.4 \pm 3.3$ & -- & $6.19 \pm 0.45$ & $4.58 \pm 0.33$ \\
& $T_{\rm eq,\,A_B=0}$ (K) & $693 \pm 18$ & $723 \pm 13$ & $421 \pm 11$ & $439 \pm 8$ & $407.2 \pm 7.4$ \\
& Eccentricity, $e$ & $0.15 \pm 0.11$ & $<0.25^{c}$ & $0.16 \pm 0.11$ & $0$ (fixed) & $0$ (fixed) \\
\bottomrule
\end{tabular}
\begin{tablenotes}
\item[$^{a}$] The masses of GJ~3090\,c and the 12.7-day candidate are minimum masses ($M\sin i$) due to the unknown inclination.
\item[$^{b}$] We assume coplanar orbits for the non-transiting planets. 
\item[$^{c}$] 95 \% upper limit from model with planetary configuration $K_b$, $C_c$, $C_d$. This model is disfavoured by the evidence by  $\Delta \ln Z=-6.8$ compared to model b4.
\end{tablenotes}
\end{threeparttable}
\end{table*}

\subsection{Assessing GJ~3090\,c's proximity to the stellar rotation period}\label{sec:proximity to stellar rotation}

The newly confirmed planet GJ~3090\,c has an orbital period of $15.9407 \pm 0.0058 $ days, which lies close to the star's measured rotation period of \Prot \,days. Given that M dwarfs are known to exhibit differential rotation \citep{Donati2008, Reinhold2015}, we have to consider the possibility that the 16-day signal observed in the radial velocity (RV) data is an artefact of stellar activity rather than a planetary companion. To rigorously test this hypothesis, we conducted several analyses.

We begin by verifying the consistency of the 16-day signal across instruments. Unlike Keplerian signals, stellar activity-induced radial velocity (RV) variations are known to be chromatic \citep{Reiners2010, Marchwinski2014, Trifonov2018}, meaning their amplitude depends on the observed wavelength. Therefore, if a signal appears with consistent amplitude and phase in both the redder NIRPS and the bluer HARPS datasets, this supports a planetary interpretation.

To test this, we reran the RV analysis using only the NIRPS or only the HARPS data. In each case, we removed the other instrument’s RV time series from the model in Equation~\ref{model components}, while keeping all other components—including the photometry and activity indicators—unchanged. The same prior distributions were used as in the fit to the combined datasets. 

\begin{figure*}[ht]
    \centering
    \includegraphics[width=1\textwidth]{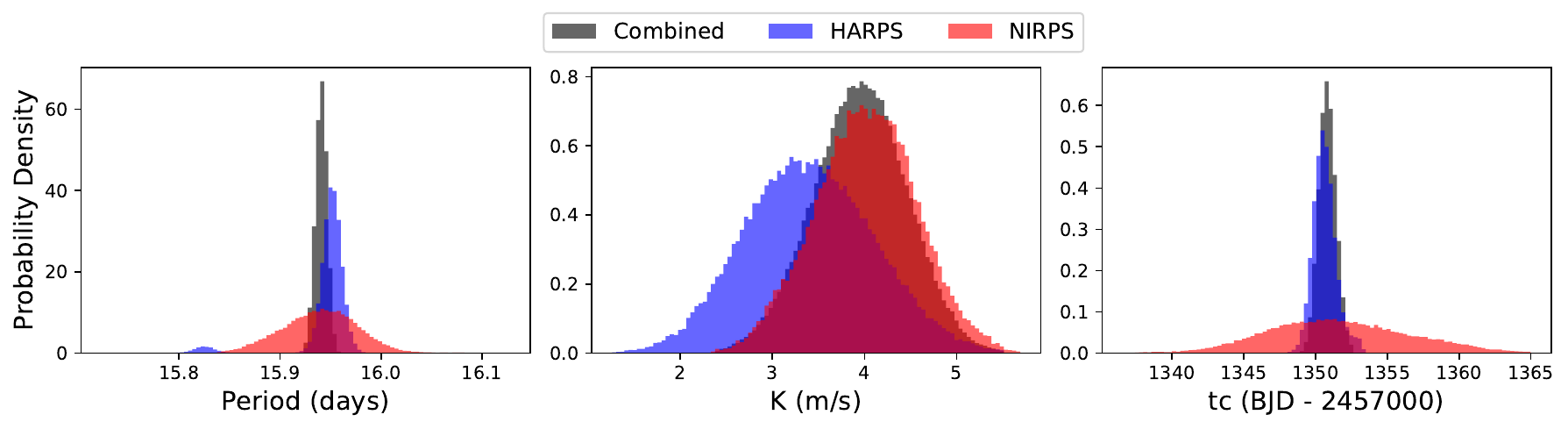}
    \caption{Posterior distributions of the Keplerian parameters of GJ~3090\,c derived from fits to the combined dataset (black), HARPS-only data (blue), and NIRPS-only data (red). The distributions show excellent agreement between HARPS and NIRPS, with tighter constraints in the combined dataset due to the increased number of RV points.}
    \label{GJ3090c_parameter_distributions}
\end{figure*}

As shown in Figure~\ref{GJ3090c_parameter_distributions}, both the HARPS-only and NIRPS-only fits yield highly consistent estimates for the orbital period, time of conjunction, and RV semi-amplitude of GJ~3090\,c. To quantify this agreement, we compute the normalized difference between the two posteriors:

\begin{equation}
    \frac{|x_{\text{N}} - x_{\text{H}}|}{\sqrt{\sigma_{\text{N}}^2 + \sigma_{\text{H}}^2}}.
\end{equation}

\noindent Here $x$ and $\sigma$ represent the medians and $1\sigma$ uncertainties from the NIRPS (N) and HARPS (H) fits, respectively. This metric quantifies the level of agreement between two independent measurements in units of their combined uncertainty. A value close to zero indicates strong consistency, while values approaching or exceeding unity suggest significant tension. For the period, RV semi-amplitude, and time of conjunction of GJ~3090\,c, we obtain normalized differences of 0.22, 0.72, and 0.16, respectively. These values lie within the combined $1\sigma$ uncertainty, confirming the agreement between the two instruments. This indicates that the signal is achromatic and thus unlikely to originate from stellar activity. It provides strong support for the planetary nature of GJ~3090\,c. Since there have been instances of achromatic activity signal in the literature (e.g. \citealt{CortsZuleta2023}; \citealt{Larue2025}), more tests are necessary. 

A widely used diagnostic to assess the planetary nature of RV signals is to monitor the evolution of the false alarm probability (FAP) in a periodogram as additional observations are added chronologically (e.g. \citetalias{Stefanov2025}). This approach tests whether the signal persists consistently throughout the full time span of observations, rather than being concentrated in a specific epoch, an outcome that could be indicative of differential rotation.

\begin{figure}[t]
    \centering
    \includegraphics[width=\columnwidth]{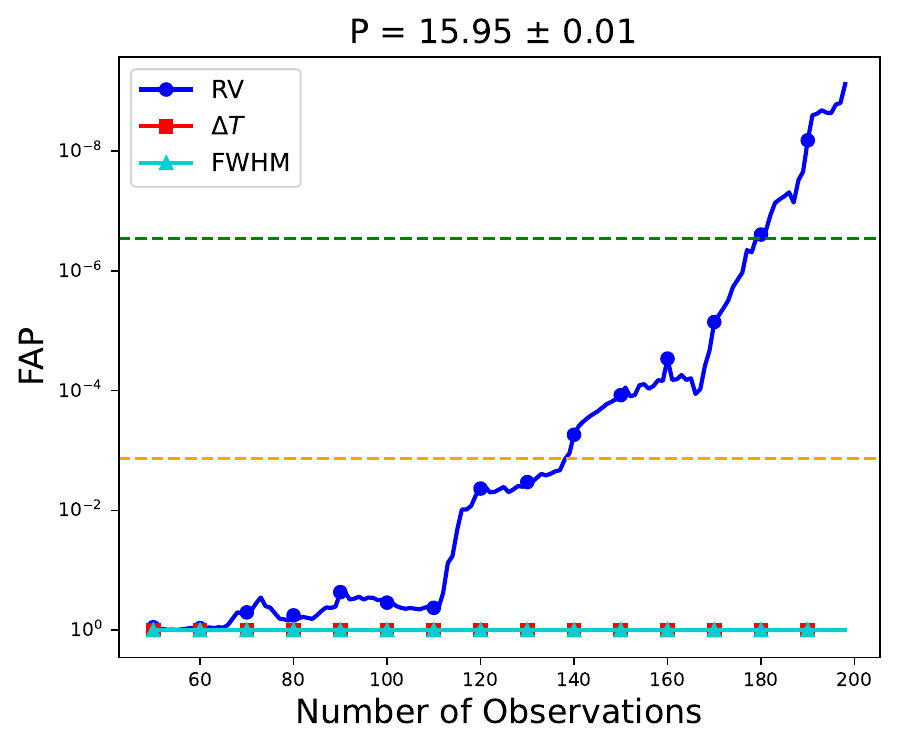}
    \caption{Chronological evolution of the false alarm probability (FAP) for the 16-day signal. As new observations are added, the FAP in the RVs decreases steadily, consistent with a stable Keplerian signal. In contrast, the FAP in the $\Delta T$ and FWHM activity indicators remains high across the datasets. The orange and green horizontal lines represent the $3\sigma$ and $5\sigma$ detection significance respectively. This behaviour supports the long-term coherence of the GJ~3090\,c signal and disfavours a stellar origin.}
    \label{fig:fap_evolution}
\end{figure}

Figure~\ref{fig:fap_evolution} shows the chronological evolution of the FAP for the 16-day signal in the RVs, compared to that in the $\Delta T$ and FWHM activity indicator, using all available HARPS (HAM and EGGS modes) and NIRPS data. As expected for a genuine Keplerian signal, the FAP associated with the RVs steadily decreases as more data are added, indicating a coherent and persistent signal across the full $\sim$6-year baseline. In contrast, the FAP of the same signal in the activity indicators remains negligible throughout the dataset, suggesting that the 16-day signal does not originate from photospheric activity. These results support the long-term dynamical stability of the GJ~3090\,c signal.

To further assess the nature of the 16-day signal, we examine how its removal affects the correlation between HARPS (HAM) radial velocities and its differential temperature indicator $\Delta T$. Because our RV model includes both first- and second-order activity terms, we analyse correlations with $\Delta T$ and its time derivative $d\Delta T/dt$.

To compute the derivative of the differential temperatures, we employed a \bestkernel kernel to model the $\Delta T$ time series from the HARPS (HAM) observations using the best-fit hyperparameters listed in Table \ref{tab:priors_posteriors}. We then calculated the time derivative of the GP prediction and interpolated these values onto the observation times while computing their uncertainties by sampling the posterior of the GP parameters and recomputing the predictions.

Before assessing these correlations, an important step is required. When plotting two out-of-phase periodic signals of the same period with different amplitudes on the y and x axis (in our case RVs and $\Delta T$) of a graph, the emergent shape is an ellipse, similar to what was observed for the correlation between RV and FWHM for Proxima and GJ~526 \citep{SuarezMascareno2020,stefanov2025_gj526}. To make the shapes linear, it is necessary to first subtract the best linear fit of each correlation (RVs vs $\Delta T$ and RVs vs $d \Delta T/dt$) to the RVs of the opposing graph. This effectively isolates the first- and second-order contributions. 

Subsequently, we repeated the linear fits after subtracting the best-fit Keplerian model for the 16-day signal from the RV data. If the signal is of stellar origin, the correlation between the RV residuals and the activity indicators ($\Delta T$ and $d \Delta T/dt$) should weaken, as part of the activity signal which causes these correlations would have been removed from the RV data. Conversely, if the signal corresponds to a planetary origin, the correlations should strengthen, since we would have removed a signal to which the indicators are blind, isolating the remaining activity-driven variations.

To quantitatively assess the correlations, we used two metrics: the Pearson correlation coefficient ($R$) and the statistical significance of the linear slope, expressed as the number of standard deviations from zero slope ($N_{\sigma}$). The results are shown in Figure \ref{fig:correlation_rv_dtemp}. Notably, both correlations improve after removing the 16-day signal, with the second-order correlation (between RV and $d \Delta T/dt$) showing a more significant improvement: $\Delta R_{\mathrm{Pearson}} = 0.10$ and $\Delta N_{\sigma} = 3.4$. This increase in correlation strength indicates that the 16-day signal is unlikely to be caused by stellar activity and more likely originates from a planetary companion.

\subsection{Comparison of GP frameworks}

In this section, we aim to compare the performance of the serial, semi-shared, and multidimensional GP models and assess the sensitivity of each model to the 15.9-day planetary signal using an injection-recovery analysis. Specifically, which models are more prone to overfitting coherent signals, regardless of their origin or period, by absorbing them into the GP component and potentially misinterpreting them as stellar variability. To test this, we construct a synthetic dataset containing only stellar activity, simulated using a quasi-periodic kernel from the \texttt{george} package \citep{george}, and add white noise matching the median noise level of the TESS, HARPS, and NIRPS datasets. The time sampling is identical to the real datasets, and we ensure that the root-mean-square (RMS) scatter matches that of the four observed time series.

We then inject circular Keplerian signals with varying periods and semi-amplitudes into this synthetic dataset. For each injection, we fit the data using our \texttt{S+LEAF} model and compute the Lomb-Scargle periodogram of the residuals. This process is repeated across a grid of periods from 1 to 1000 days and semi-amplitudes from 0.5 to 10~m\,s$^{-1}$. The resulting false alarm probabilities (FAPs) at each grid point are converted into detection significances using the standard inverse Gaussian cumulative distribution function (CDF),

\begin{equation}
    \sigma = \Phi^{-1}(1 - \mathrm{FAP}),
\end{equation}

\noindent where $\Phi^{-1}$ is the quantile function (CDF$^{-1}$) of the standard normal distribution.

\begin{figure*}[ht]
    \centering
    \includegraphics[width=\textwidth]{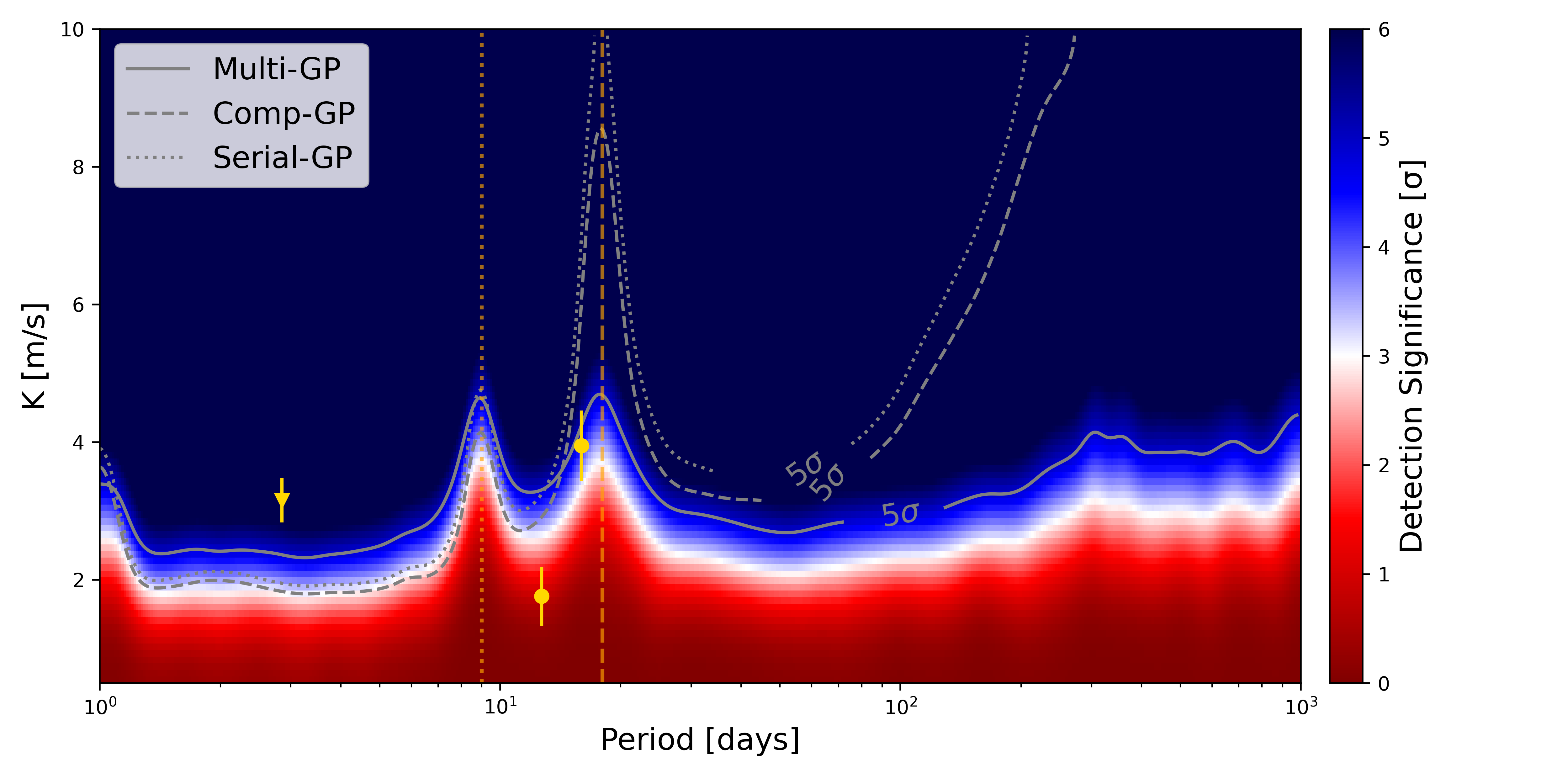}
    \caption{Sensitivity map showing the detection significance ($\sigma$) of the multidimensional GP approach as a function of period and semi-amplitude $K$ of injected planets. The colour map indicates the detection significance computed from the periodogram of the residuals. The 5$\sigma$ detection contours are shown for the serial GP (dotted grey line), the composite or semi-shared GP (dashed grey line), and the multidimensional GP (solid grey line). Known planets are marked with yellow symbols. The vertical orange lines indicate the stellar rotation period and its first harmonic. The 16-day signal is recovered with a detection significance of $4.05\,\sigma$ using the multidimensional GP, $1.10 \, \sigma$ using the composite GP and $0\,\sigma$ with the serial GP.}
    \label{fig:sensitivity_map}
\end{figure*}

The resulting sensitivity map is shown in Figure~\ref{fig:sensitivity_map}, where the detection significance is encoded in the colour scale. Known planets from our dataset are overlaid with yellow markers, and vertical orange lines indicate the stellar rotation period ($P_\mathrm{rot}$) and its first harmonic.

We also compare this multidimensional GP model to two other types of model which inform stellar activity from activity indicators in different ways: The serial model and the composite or semi-shared model. In the serial model, GP hyperparameters are inferred from activity indicators and their posteriors are used as priors to fit the RVs. In the composite or semi-shared model, each time serie has its own covariance matrix, but some GP hyperparameters are fitted simultaneously between all datasets. In our case the rotation period and the length scale of the GPs are shared.  

As illustrated in Figure~\ref{fig:sensitivity_map}, the multidimensional GP model exhibits greater sensitivity than the serial GP across multiple regions of parameter space. It is the case near the stellar rotation period ($\sim 18$d), where activity-induced signals dominate. The multidimensional approach also shows a greater sensitivity at long periods where the detection region of the serial and composite GP models is restricted by overfitting. This means that the multidimensional model should be prioritized when looking for long period planets or magnetic cycles. In our case, the composite method seems to give better detection significance for small periods as well as near the first harmonic of the rotation period.

This injection test informs us that the 16-day signal corresponding to GJ~3090\,c is recovered with a significance of $4.05\,\sigma$ using our multidimensional GP framework, whereas it is detected with lower significance with other methods: $1.10\,\sigma$ when using the composite GP approach and a null detection ($0 \, \sigma$) for the serial GP. While the recovered significance is only $\sim4\sigma$, this detection map is based on a false alarm probability (FAP) statistic, which itself relies on the analytical approximation from \citet{Baluev2008}. As such, it should not be directly compared with the detection significance derived from Bayesian model comparison using the log-evidence difference ($\Delta \ln Z$). Some variation between the $\sigma$ detection inferred from a FAP and that from $\Delta \ln Z$ is therefore expected. In our study, this diagnostic primarily serves to illustrate the relative performance of the different GP approaches. Nevertheless, this exercise strengthens our conclusion that the multidimensional GP model is the most effective approach to prevent overfitting the planetary signal. Instead, it effectively distinguishes the planetary nature of GJ~3090\,c from stellar activity. The lower sensitivity of the serial GP method, combined with the more limited dataset used in \citetalias{AlmenaraBonfils2022}, may explain why this signal was not previously reported.

We note that the same analysis using the actual residuals from the data (after subtracting known planetary and magnetic cycle signals) has been tried, and the results were consistent with those from the synthetic dataset. We chose to show the results from the latter to avoid contamination from unmodelled signals in the real data (e.g. low-amplitude planets).

\subsection{Stability and resonant analysis}

We analyse the stability of the GJ 3090 system with the additions of both GJ 3090 c and the 12.7-day candidate and analyse the system's resonant state using the planetary parameters derived in this work. We first perform an introductory simulation to determine whether or not this system is potentially stable. To assess the state of the system, we run a numerical simulation using Rebound and the default IAS15 integrator \citep{Rein2012}. We initialize GJ 3090 and all three planets or planet candidates using the median values of planetary masses, periods, eccentricities, and epochs given in Table \ref{tab:planet_parameters}. As GJ 3090 c and the 12.7-day candidate are not transiting, we do not know their inclinations relative to GJ 3090 b; as such, we assume all three orbits are coplanar. We integrate the system for 1 Myr, saving planetary parameters for all three bodies every 1000 years.

We find that all three planets remain stable for the entirety of the 1 Myr integration, with periods for all three planets remaining nearly unchanged and eccentricities remaining very low ($<0.02$) for the whole integration. While further analysis of the stability of the GJ 3090 system is beyond the scope of this paper, it appears that the GJ 3090 system is not immediately unstable with the addition of this planet candidate.

In addition, we note that GJ 3090 c and the 12.7-day candidate have a period ratio very close to $5:4$, potentially suggestive of the planet being in a two-body mean motion resonance (MMR). As such, we attempt to identify if such a resonance exists between these two planets.

Two-body MMRs exist when a pair of planets have librating resonance angles $\phi_1, \phi_2$, given by

\begin{align}
\phi_1 & = p\lambda_i-q\lambda_j-(p-q)\omega_i \,, \\    
\phi_2 & = p\lambda_i-q\lambda_j-(p-q)\omega_j \,,  
\end{align}

where $\lambda$ is the standard mean longitude of the planet and $\omega$ is the argument of periapsis of the planet, while the period ratio of the two planets is $p/q$ for integers $p, q$ (in our case, $5$ and $4$, respectively). If these two angles librate rather than circulate, the two planets are said to be in an MMR. Alternatively, one can use the composite two-body resonance angle $\phi_{12}$, given by \citet{Laune2022} to be

\begin{equation}
\tan \phi_{12} = \frac{f_1e_1\sin\phi_1+f_2e_2\sin\phi_2}{f_1e_1\cos\phi_1+f_2e_2\cos\phi_2}\,,
\end{equation}

where $e$ is the eccentricity of a planet and $f_1, f_2$ are coefficients that depend on the values of $p, q$ (see Table 1 of \citealp{Deck2013}). Planets can still be in resonance even if $\phi_1, \phi_2$ circulate, so long as the composite two-body resonance angle still librates.

\begin{figure}[h]
    \centering
    \includegraphics[width=\linewidth]{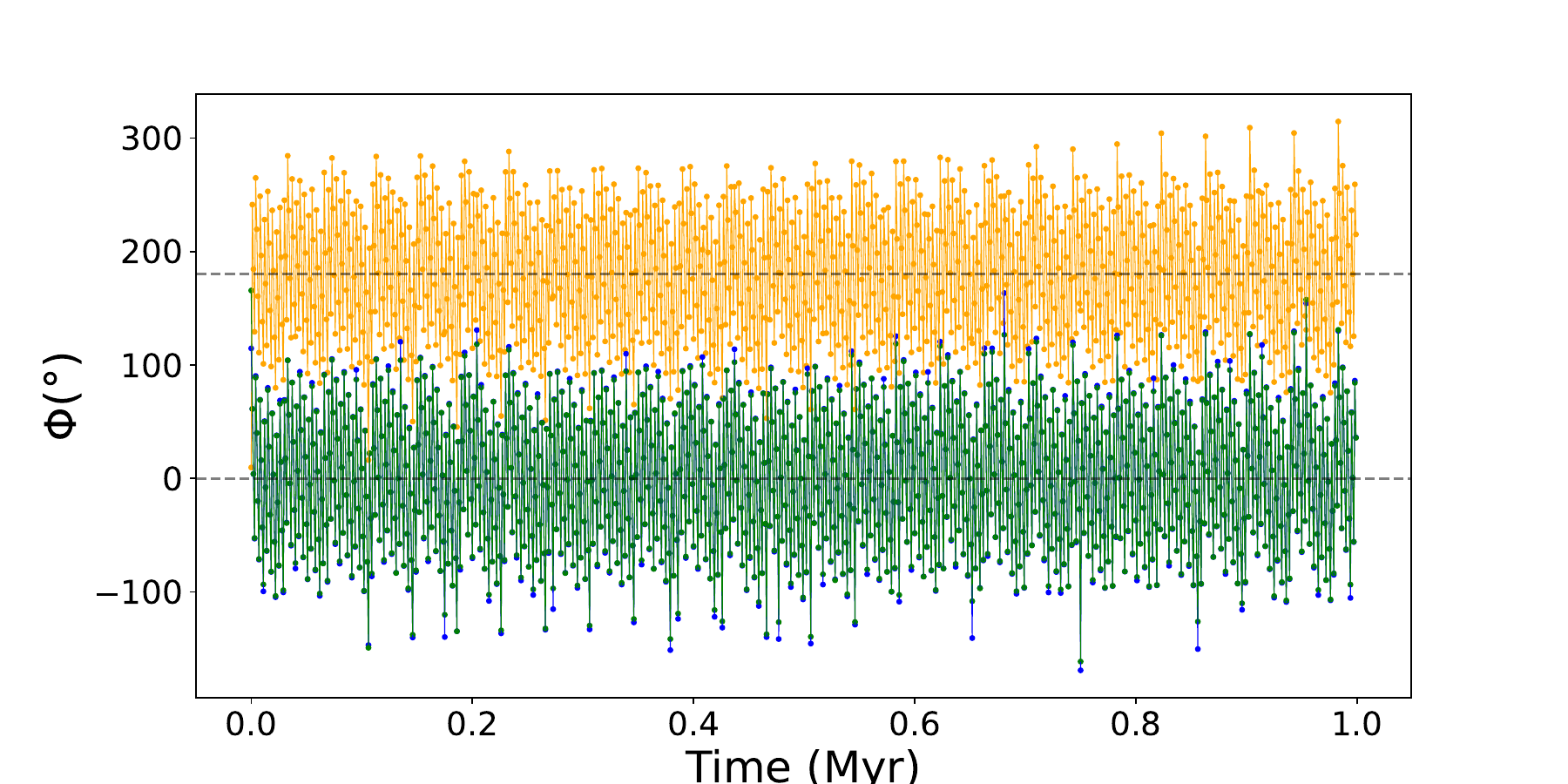}
    \caption{Plot of samples of $\phi_1$ (orange), $\phi_2$ (blue), and $\phi_{12}$ (green) between GJ 3090 c and the 12.7-day candidate. Samples are taken every 1000 years, across a 1 Myr integration of the GJ 3090 system. Horizontal dashed lines corresponding to $0^\circ$ and $180^\circ$ are shown as well. All three angles appear to librate, though this is merely an artefact of the sampling frequency.}
    \label{fig:REBOUND-1Myr}
\end{figure}

\begin{figure}[h]
    \centering
    \includegraphics[width=\linewidth]{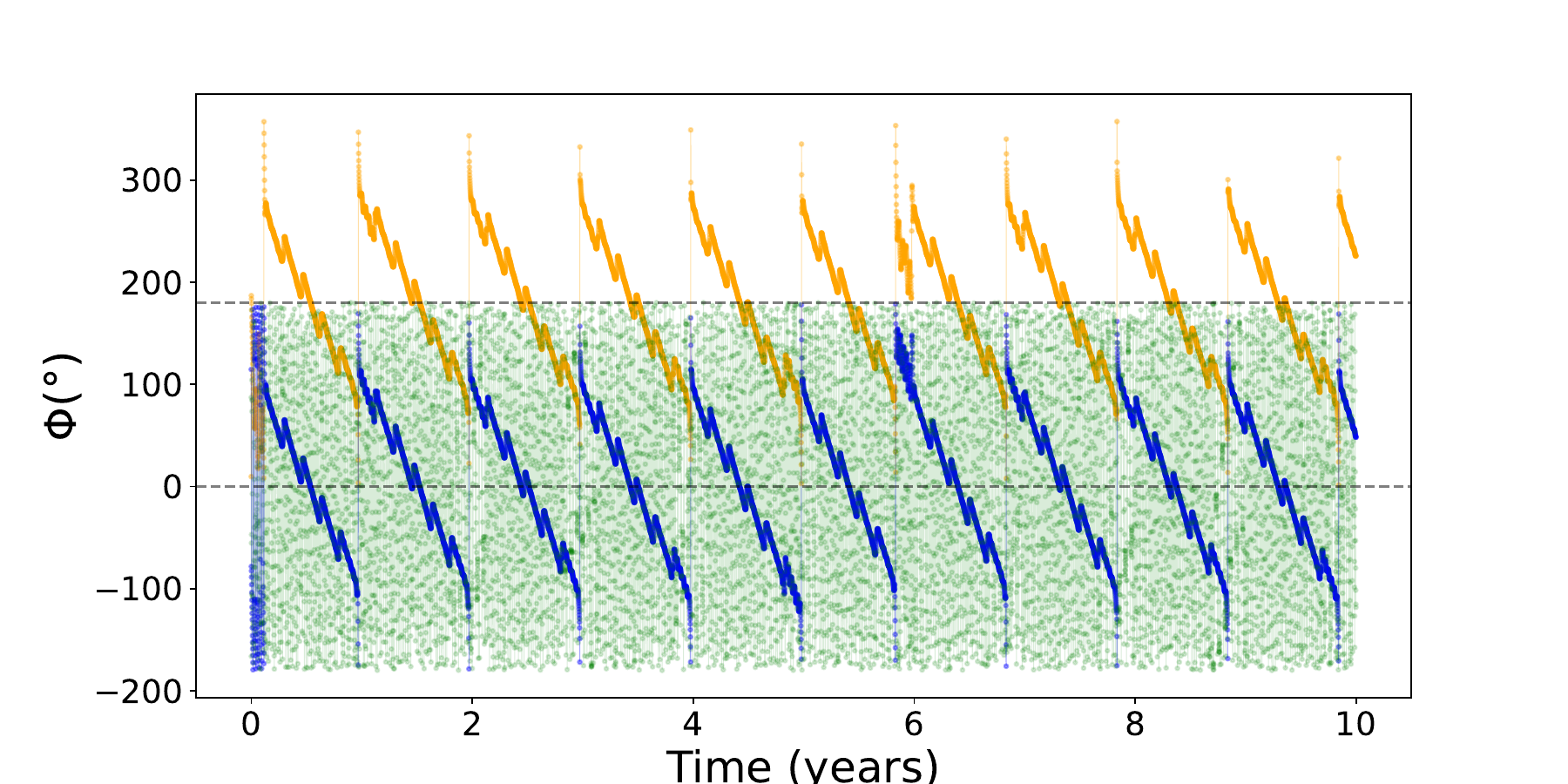}
    \caption{Plot of samples of $\phi_1$ (orange), $\phi_2$ (blue), and $\phi_{12}$ (green) between GJ 3090 c and the 12.7-day candidate. Samples are taken 1000 times a year, across the first 10 years of a 1 Myr integration of the GJ 3090 system. Horizontal dashed lines corresponding to $0^\circ$ and $180^\circ$  are shown as well. While Figure \ref{fig:REBOUND-1Myr} appears to show all three angles librating, the higher sampling frequency shown here shows that all three angles circulate.}
    \label{fig:REBOUND-10yr}
\end{figure}

We calculate all three of these angles every 1000 years across the entirety of the integration. We additionally increase the sampling rate drastically for the first 10 years of the integration, sampling 1000 times per year for only this period. The results are shown in Figures \ref{fig:REBOUND-1Myr} and \ref{fig:REBOUND-10yr}, respectively. 

As can be seen in Figure \ref{fig:REBOUND-1Myr}, all three resonant angles appear to be librating around $180^\circ$, $0^\circ$, and $0^\circ$, respectively, which would be indicative of GJ 3090 c and the 12.7-day candidate indeed being in a $5:4$ two-body MMR. However, as can be seen in Figure \ref{fig:REBOUND-10yr}, a higher sampling frequency shows a very different result. The angles $\phi_1$ and $\phi_2$ appear to be circulating, just with a circulation frequency very near to $1$ year. Meanwhile, $\phi_{12}$ appears to be circulating much faster. While this integration shows that the system is not in resonance, this system appears to be very nearly resonant, with the slow circulation of $\phi_1$ and $\phi_2$ that takes many orbits to circulate potentially even indicating nodding in and out of resonance \citep{Ketchum2013}. As the resonant state of a system can depend on very slight changes in planet mass or period, however, more precise determination of planetary parameters is necessary to better determine the resonant state of the system.

\subsection{Updated parameters of the GJ~3090 system}
We update the stellar rotation period of the host star GJ~3090 to \Prot \, days, a small increase from the previous estimate of $17.729^{+0.17}_{-0.036}$ days reported by \citetalias{AlmenaraBonfils2022}. This $1.3\,\sigma$ shift may be explained by the presence of the 16-day planetary signal associated with GJ~3090\,c, which was most likely absorbed by the GP activity model used in the previous study. We recover both the 18-day rotational period and its first harmonic from the RV data after removing all planetary contributions, as shown in the periodogram in Figure~\ref{fig:periodograms}. The forest of peaks near the rotation period in the periodogram shows that the star most likely experienced differential rotation during the $\sim~6$\,yr timespan of our dataset.

We provide refined mass and radius measurements for GJ~3090\,b of \massb \, $M_{\oplus}$ and \radius \, $R_{\oplus}$ respectively, and confirm the existence of a second planet, GJ~3090\,c, orbiting the star with a period of $15.94$ days and a mass of \massc \, $M_{\oplus}$. We also detect a 12.7-day planetary candidate originally reported by \citetalias{AlmenaraBonfils2022}, now with an updated mass of \masscandidateone \, $M_{\oplus}$.

\begin{figure}[ht]
    \centering
    \hspace{-0.8cm}
    \includegraphics[width=\columnwidth]{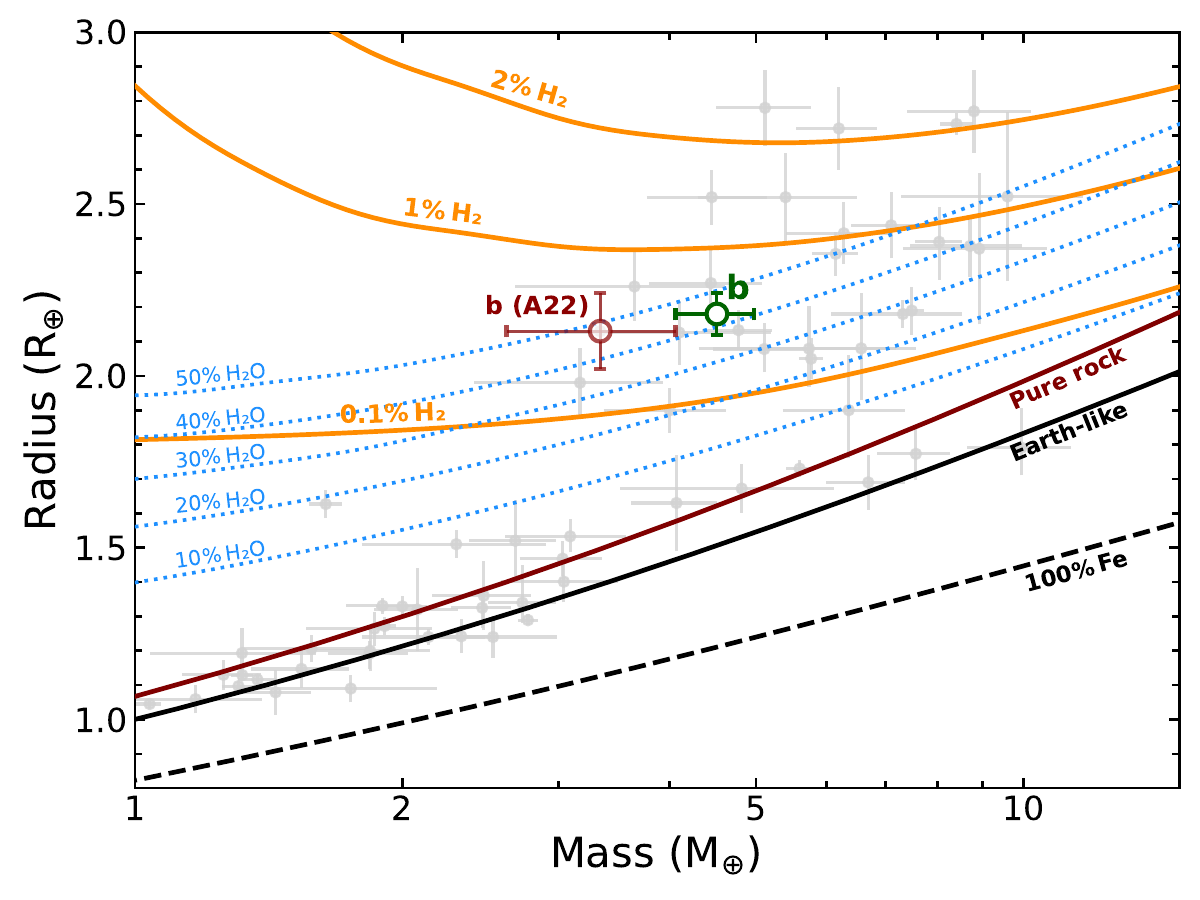}
    \caption{Mass–radius diagram. GJ~3090\,b (green) is shown alongside known M  dwarf exoplanets with $<30$\% uncertainty in both mass and radius (grey points). The red symbol indicates the mass-radius position of GJ~3090\,b as previously measured by \citetalias{AlmenaraBonfils2022}. Theoretical iso-composition lines include purely rocky compositions (black and red), H$_2$O envelope (steam), planets with Earth-like rocky interior (dotted blue), and H/He envelope sub-Neptunes with Earth-like rocky interior (orange). These interiors are retrieved using the detailed modelling described in Sect.~\ref{sec:interior} and assume $T_{eq}=700$\,K with $1$ Gyr system age.}
    \label{fig:mass_radius_diagram}
\end{figure}

GJ~3090\,b is placed in the mass–radius (M–R) diagram shown in Figure~\ref{fig:mass_radius_diagram}. This figure includes known exoplanets orbiting M dwarfs with better than 30\% precision in both mass and radius (grey points). We overlay theoretical M–R relations from several interior models: from pure iron to pure silicate compositions \citep{Zeng2019}, water-rich envelopes \citep{Aguichine2021}, and H/He envelopes \citep{Lopez2014}. The revised mass of GJ~3090\,b increases its density and brings it closer to the bulk population of sub-Neptunes around M dwarfs. Its position in the M–R diagram lies within a region of degeneracy between water-rich and H/He-dominated compositions, suggesting a range of plausible internal structures.

In terms of incident stellar flux, all three planets receive significantly more radiation than Earth, as summarized in Table~\ref{tab:planet_parameters}. This places them well outside the habitable zone of the system, which is estimated to lie around orbital periods of $\sim100$ days. Finally, while we cannot completely rule out the existence of additional planetary companions in the system, no statistically significant peaks are observed in the periodograms of the RV residuals after modelling the three identified planets and stellar activity (Figure~\ref{fig:periodograms}).

\subsection{Detailed interior modelling of GJ~3090~b}\label{sec:interior}

We investigate the composition of GJ~3090~b using detailed interior structure modelling coupled to the \texttt{emcee} affine-invariant ensemble sampler \citep{Foreman-Mackey_2013} to characterize the probability density of possible atmospheric mass fraction (AMF) and core mass fraction (CMF). The planetary interior is assumed to be composed of an Fe-Ni metallic core and a silicate mantle (\texttt{SUPEREARTH} \citealt{Valencia_2007}), while the outermost layer consists of either a hydrogen-helium envelope or a water vapour atmosphere modelled using the CEPAM code \citep{Guillot_1995}, with equations of state from \citet{Saumon_1995} for H/He and \citet{French_2009} for H$_2$O. In all cases, we assume that the rocky interior contains no volatiles and follow the numerical set-up given in \citet{Plotnykov_2020}.

We consider two sub-Neptune composition scenarios: (1) the planet has a H/He envelope (75\% H$_2$ to 25\% He) and (2) the planet has a pure H$_2$O envelope. For these scenarios, we impose stellar-informed priors on the rocky interior based on the host star’s refractory abundances taken from Table~\ref{tab:stellar_abundances}:

\begin{align*}
 \mathrm{Fe/Mg_{planet}} &\sim \mathcal{N}(\mathrm{Fe/Mg_{star}},\sigma_{\rm star}^2)\,, \\
 \mathrm{Fe/Si_{planet}} &\sim \mathcal{N}(\mathrm{Fe/Si_{star}},\sigma_{\rm star}^2)
\end{align*}

where all ratios are by weight. Additionally, the mantle mineralogy is allowed to vary in terms of the Bridgmanite to Wustite ratio (MgSiO$_3$ vs MgO, $x_{\mathrm{Wu}}$). These assumptions effectively constrain the rocky core mass fraction ($\mathrm{CMF}=\frac{\mathrm{rcmf}}{\mathrm{rcmf}+1}$, where rcmf is the core-to-mantle mass ratio, $M_{\mathrm{core}}/M_{\mathrm{mantle}}$) of the planet and mitigate the problem of compositional degeneracy.

Considering case (1), where GJ~3090~b has retained its primordial H/He envelope, we recover AMF~$=0.006 \pm 0.002$ and CMF~$=0.35^{+0.04}_{-0.05}$ when using stellar-informed priors (Appendix~\ref{fig:corner_H2He_priors}). If we impose no prior on the rocky composition, the AMF increases slightly to $0.007^{+0.06}_{-0.04}$, while the CMF distribution becomes much broader, with CMF~$=0.5 \pm 0.3$ (also shown in Appendix~\ref{fig:corner_H2He_priors}). For case (2), where the envelope is composed of pure water vapour, we find WMF~$=0.43^{+0.08}_{-0.07}$ and CMF~$=0.35^{+0.04}_{-0.05}$, with the posterior distributions shown in Appendix~\ref{fig:corner_H2O_priors}.

\subsection{Linking interior structure and atmospheric studies of GJ 3090b}

The results of our interior structure modelling indicate that both the AMF and the WMF should be considered as upper limits. This is because degeneracies in the mass-radius relationship prevent a definitive separation between a purely H/He-dominated or water-dominated envelope, and the planet is more likely to possess a mixed composition with both components present. In particular, the constraint that AMF is below 1\% is well under the predictions of atmosphere formation models, which typically expect initial H/He envelopes of at least a few percent by mass for sub-Neptunes at this size and insolation \citep{Ginzburg2016, Venturini2017, Rogers2023}. Such a low H/He fraction would require either formation in a gas-poor environment \citep{Lee2021}, rapid dissipation of the gas disk \citep{Ercolano2017}, or significant loss through atmospheric escape or impacts \citep{Inamdar2016}.

High-resolution CRIRES+ spectra of GJ~3090\,b \citep{Parker2025} revealed no signs of molecular species, consistent with either a highly metal-enriched atmosphere with mean molecular weight $> 7.1$~g~mol$^{-1}$ and H/He fraction $<33$\%, or the presence of high-altitude aerosols at pressures below $3.3\times10^{-5}$~bar that obscure molecular features. Similarly, JWST observations \citepalias{Ahrer2025} show highly muted spectral features and a lack of CH$_4$. A key result from the JWST study is the detection of metastable helium absorption at 10833~\AA\ at $5.5\sigma$ significance, representing the first such detection in a sub-Neptune with JWST. This demonstrates the presence of a H/He atmosphere, even if its overall mass fraction is small. The measured helium signal is an order of magnitude weaker than solar-metallicity models predict, and the most plausible explanation is that the atmosphere is highly enriched in heavy elements, which would reduce both the mass-loss rate and the amplitude of the helium feature. Based on hydrodynamic modelling, \citetalias{Ahrer2025} estimate that photoevaporation would require $\sim50$~Gyr to strip the atmosphere entirely, implying that the current envelope is long-lived.

The combination of an H/He detection, the lack of molecular features in both CRIRES+ and JWST spectra, and the high inferred metallicity strongly supports the presence of a substantial water component mixed with H/He, making GJ~3090\,b a compelling water-world candidate. This interpretation is reminiscent of the atmospheric properties inferred for K2-18\,b \citep{Benneke2019, Madhusudhan2020} and TOI-270\,d \citep{Holmberg2024}, both of which exhibit evidence for volatile-rich envelopes with high metallicities. In the case of GJ~3090\,b, the atmospheric constraints thus provide a coherent picture with the interior structure modelling, pointing towards a volatile-rich planet where water likely constitutes a significant fraction of the envelope.

\section{Summary and conclusions} \label{sec:conclusion}

We presented an updated characterization of the planetary system orbiting the nearby M2 dwarf GJ~3090 using radial velocity data from HARPS and NIRPS, combined with TESS photometry and the $\Delta T$ spectroscopic activity indicator. Using a multidimensional GP framework tailored to simultaneously model stellar activity across multiple datasets, we derived new mass constraints and provided a more robust interpretation of the system's architecture.

Our analysis significantly improves the mass measurement of the transiting planet GJ~3090\,b, now determined to be \massb \, $M_{\oplus}$. Combined with a radius of \radius \, $R_{\oplus}$ we derived from four TESS sectors, this yields a new density estimate of \density \,g\,cm$^{-3}$, suggesting a volatile-rich composition. It places the planet in a highly degenerate section of the M-R diagram that can be reproduced by different ratios of H$_2$O and H/He, and brings it closer to the bulk distribution of sub-Neptunes around M dwarfs.

We confirm the existence of a second planet, GJ~3090\,c, with a 15.94-day orbit and a minimum mass of \massc \, $M_{\oplus}$. Although the signal lies close to the stellar rotation period ($P_\mathrm{rot} \approx 18$\,days), we provided multiple lines of evidence supporting its planetary nature. These include consistency between the HARPS and NIRPS RV signals, the inability of our GP activity model to absorb the 16-day signal in injection-recovery tests, its robust detection across model comparisons using different GP kernels, and the improved correlation between the HARPS RVs and the $\Delta T$ activity indicator once the 16-day planetary signal is subtracted.

We also report evidence for a third planet candidate at 12.7 days with a minimum mass of \masscandidateone \, $M_\oplus$, initially proposed by \citet{AlmenaraBonfils2022}. While this signal does not reach the same level of statistical significance as GJ~3090\,c, it remains a compelling target for continued monitoring as planets of this mass typically lie in the radius valley and could therefore provide insight into its origin.

Importantly, our study illustrates the advantages of multidimensional GP approaches in disentangling stellar activity from planetary signals, particularly for signals near the stellar rotation period. Compared to traditional sequential or semi-shared GP methods, our framework mitigates the risk of overfitting and increases sensitivity to both signals close to the rotation period and long-period signals.

The GJ~3090 system thus emerges as a compact multi-planetary system, possibly composed of three sub-Neptunes, with only the innermost planet observed in transit. Assuming coplanar orbits, we explain the absence of transits for the outer companions through geometric arguments based on their derived impact parameters. No additional significant signals are found in the residuals, but the system remains an interesting candidate for further RV monitoring.

Future observations, including further RV measurements and photometric campaigns, will help confirm the nature of the 12.7-day candidate and probe the presence of additional companions in the system. GJ~3090\,b remains a particularly attractive target for atmospheric and interior studies, with its mass and radius now measured to 10\% precision or lower.

\begin{acknowledgements}
PLam, CC, AL, LMo, RD, FBa, BB, LMa, RA, LB, \'EA, AB, AD-B, LD, NJC, OL, JS-A, PV, TV \& JPW  acknowledge the financial support of the FRQ-NT through the Centre de recherche en astrophysique du Qu\'ebec as well as the support from the Trottier Family Foundation and the Trottier Institute for Research on Exoplanets.
AL  acknowledges support from the Fonds de recherche du Qu\'ebec (FRQ) - Secteur Nature et technologies under file \#349961.
This work has been carried out within the framework of the NCCR PlanetS supported by the Swiss National Science Foundation under grants 51NF40\_182901 and 51NF40\_205606.
AKS, JIGH, RR, ASM, FGT, NN, VMP \& JLR  acknowledge financial support from the Spanish Ministry of Science, Innovation and Universities (MICIU) projects PID2020-117493GB-I00 and PID2023-149982NB-I00.
AKS  acknowledges financial support from La Caixa Foundation (ID 100010434) under the grant LCF/BQ/DI23/11990071.
LMo  acknowledges the support of the Natural Sciences and Engineering Research Council of Canada (NSERC), [funding reference number 589653].
RD, FBa, LMa, TA, \'EA, J-SM, MO, JS-A \& PV  acknowledges support from Canada Foundation for Innovation (CFI) program, the Universit\'e de Montr\'eal and Universit\'e Laval, the Canada Economic Development (CED) program and the Ministere of Economy, Innovation and Energy (MEIE).
SCB, ED-M, NCS, EC, ARCS \& JGd  acknowledge the support from FCT - Funda\c{c}\~ao para a Ci\^encia e a Tecnologia through national funds by these grants: UIDB/04434/2020, UIDP/04434/2020.
SCB   acknowledges the support from Funda\c{c}\~ao para a Ci\^encia e Tecnologia (FCT) in the form of a work contract through the Scientific Employment Incentive program with reference 2023.06687.CEECIND and DOI \href{https://doi.org/10.54499/2023.06687.CEECIND/CP2839/CT0002}{10.54499/2023.06687.CEECIND/CP2839/CT0002.}
XB, XDe, ACar, TF \& VY  acknowledge funding from the French ANR under contract number ANR\-24\-CE49\-3397 (ORVET), and the French National Research Agency in the framework of the Investissements d'Avenir program (ANR-15-IDEX-02), through the funding of the ``Origin of Life" project of the Grenoble-Alpes University.
The Board of Observational and Instrumental Astronomy (NAOS) at the Federal University of Rio Grande do Norte's research activities are supported by continuous grants from the Brazilian funding agency CNPq. This study was partially funded by the Coordena\c{c}\~ao de Aperfei\c{c}oamento de Pessoal de N\'ivel Superior—Brasil (CAPES) — Finance Code 001 and the CAPES-Print program.
BLCM \& AMM  acknowledge CAPES postdoctoral fellowships.
BLCM  acknowledges CNPq research fellowships (Grant No. 305804/2022-7).
NBC  acknowledges support from an NSERC Discovery Grant, a Canada Research Chair, and an Arthur B. McDonald Fellowship, and thanks the Trottier Space Institute for its financial support and dynamic intellectual environment.
DBF  acknowledges financial support from the Brazilian agency CNPq-PQ (Grant No. 305566/2021-0). Continuous grants from the Brazilian agency CNPq support the STELLAR TEAM of the Federal University of Ceara's research activities.
JRM  acknowledges CNPq research fellowships (Grant No. 308928/2019-9).
ED-M  further acknowledges the support from FCT through Stimulus FCT contract 2021.01294.CEECIND. ED-M  acknowledges the support by the Ram\'on y Cajal contract RyC2022-035854-I funded by MICIU/AEI/10.13039/501100011033 and by ESF+.
XDu  acknowledges the support from the European Research Council (ERC) under the European Union’s Horizon 2020 research and innovation programme (grant agreement SCORE No 851555) and from the Swiss National Science Foundation under the grant SPECTRE (No 200021\_215200).
DE  acknowledge support from the Swiss National Science Foundation for project 200021\_200726. The authors acknowledge the financial support of the SNSF.
ICL  acknowledges CNPq research fellowships (Grant No. 313103/2022-4).
CMo  acknowledges the funding from the Swiss National Science Foundation under grant 200021\_204847 “PlanetsInTime”.
Co-funded by the European Union (ERC, FIERCE, 101052347). Views and opinions expressed are however those of the author(s) only and do not necessarily reflect those of the European Union or the European Research Council. Neither the European Union nor the granting authority can be held responsible for them.
GAW is supported by a Discovery Grant from the Natural Sciences and Engineering Research Council (NSERC) of Canada.
0
KAM  acknowledges support from the Swiss National Science Foundation (SNSF) under the Postdoc Mobility grant P500PT\_230225.
RA  acknowledges the Swiss National Science Foundation (SNSF) support under the Post-Doc Mobility grant P500PT\_222212 and the support of the Institut Trottier de Recherche sur les Exoplan\`etes (IREx).
We acknowledge funding from the European Research Council under the ERC Grant Agreement n. 337591-ExTrA.
LB  acknowledges the support of the Natural Sciences and Engineering Research Council of Canada (NSERC).
This project has received funding from the European Research Council (ERC) under the European Union's Horizon 2020 research and innovation programme (project {\sc Spice Dune}, grant agreement No 947634). This material reflects only the authors' views and the Commission is not liable for any use that may be made of the information contained therein.
ARCS  acknowledges the support from Funda\c{c}ao para a Ci\^encia e a Tecnologia (FCT) through the fellowship 2021.07856.BD.
LD  acknowledges the support of the Natural Sciences and Engineering Research Council of Canada (NSERC) and from the Fonds de recherche du Qu\'ebec (FRQ) - Secteur Nature et technologies.
FG  acknowledges support from the Fonds de recherche du Qu\'ebec (FRQ) - Secteur Nature et technologies under file \#350366.
This work was supported by grants from eSSENCE (grant number eSSENCE@LU 9:3), the Swedish National Research Council (project number 2023 05307), The Crafoord foundation and the Royal Physiographic Society of Lund, through The Fund of the Walter Gyllenberg Foundation."
NN  acknowledges financial support by Light Bridges S.L, Las Palmas de Gran Canaria.
NN acknowledges funding from Light Bridges for the Doctoral Thesis "Habitable Earth-like planets with ESPRESSO and NIRPS", in cooperation with the Instituto de Astrof\'isica de Canarias, and the use of Indefeasible Computer Rights (ICR) being commissioned at the ASTRO POC project in the Island of Tenerife, Canary Islands (Spain). The ICR-ASTRONOMY used for his research was provided by Light Bridges in cooperation with Hewlett Packard Enterprise (HPE).
CPi  acknowledges support from the NSERC Vanier scholarship, and the Trottier Family Foundation. CPi  also acknowledges support from the E. Margaret Burbidge Prize Postdoctoral Fellowship from the Brinson Foundation.
TV  acknowledges support from the Fonds de recherche du Qu\'ebec (FRQ) - Secteur Nature et technologies under file \#320056.

We used the following \texttt{Python} packages for data analysis and visualization: \texttt{numpy} \citep{Harris2020}, \texttt{scipy} \citep{Virtanen2020}, \texttt{matplotlib} \citep{Hunter2007}, \texttt{astropy} \citep{TheAstropyCollaboration2022}, \texttt{juliet} \citep{Espinoza2019_juliet}, \texttt{spleaf} \citep{Delisle2022}, and \texttt{NIEVA} (Stefanov et al., in prep.).

This manuscript was written and compiled in \texttt{OVERLEAF}. 
\end{acknowledgements}

\bibliographystyle{aa}
\bibliography{bibliography}

%\clearpage
\appendix
\onecolumn

\section{Supplementary material}

\noindent
\begin{minipage}{\columnwidth}
    \centering
    \includegraphics[width=\columnwidth]{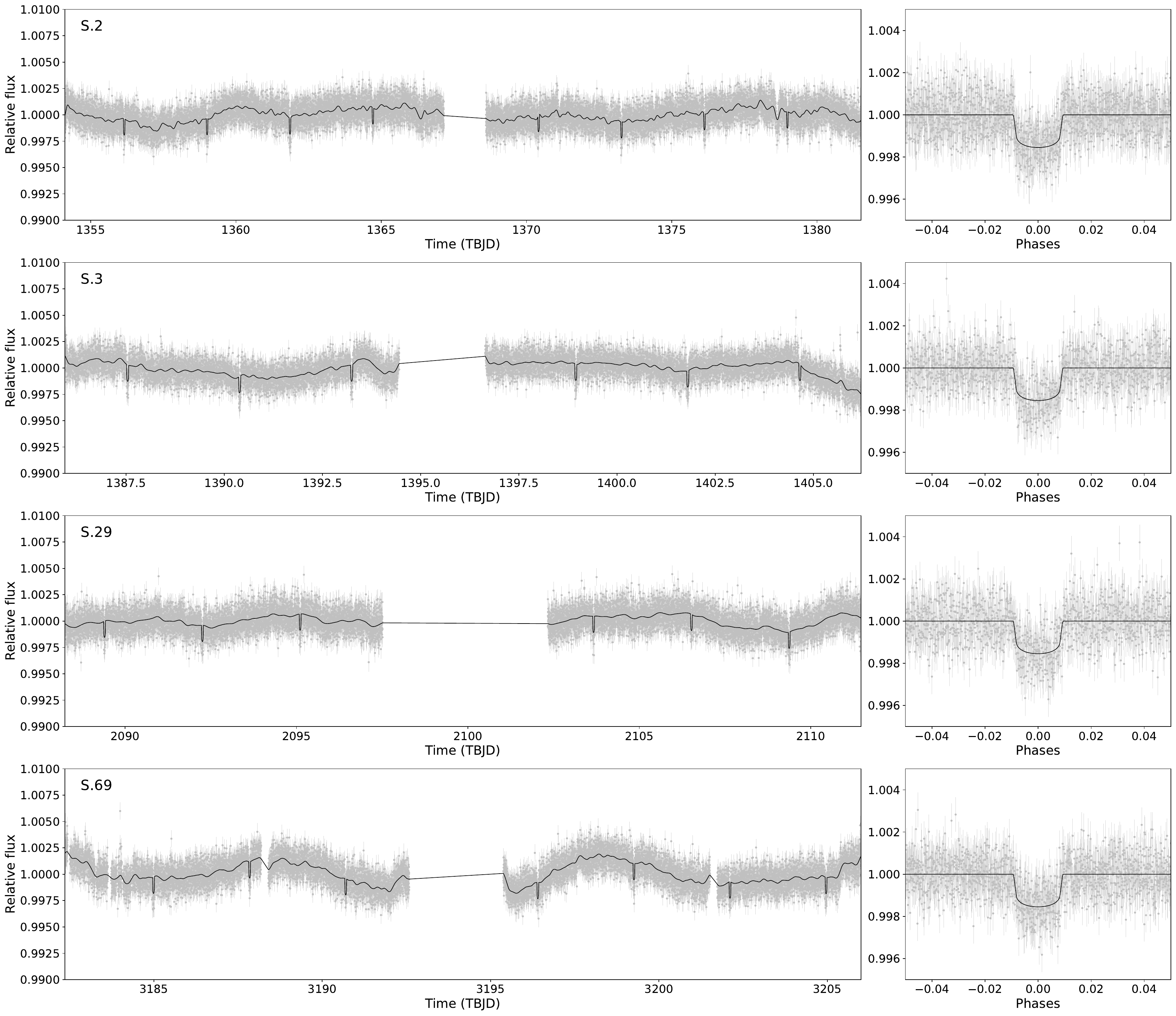}
    \captionof{figure}{TESS PDCSAP flux light curves of the four different sectors with the best-fit \texttt{juliet} model for GJ 3090~b transits shown as a black line.}
    \label{fig:full_sectors_LCs_TESS}

    \vspace{0.3cm} 

    \renewcommand{\arraystretch}{1.1}
    \small
    \captionof{table}{Posterior distributions of planetary parameters for HARPS-only, NIRPS-only, and combined fits.}
    \label{tab:planet_posteriors}
    \begin{tabular}{lccc}
    \toprule
    \textbf{Parameter} & \textbf{HARPS-only} & \textbf{NIRPS-only} & \textbf{HARPS+NIRPS} \\
    \midrule
    \multicolumn{4}{c}{\textbf{GJ~3090\,b}} \\
    $P$ [days] & $2.85310146 \pm 0.00000059$ & $2.85310191 \pm 0.00000054$ & $2.85310195 \pm 0.00000081$ \\
    $K$ [m/s] & $2.83^{+0.50}_{-0.53}$ & $3.34^{+0.25}_{-0.28}$ & $3.15 \pm 0.32$ \\
    $T_0$ [BJD-2457000] & $1356.15309 \pm 0.00019$ & $1356.15309^{+0.00017}_{-0.00019}$ & $1356.15322^{+0.00022}_{-0.00024}$ \\
    \midrule
    \multicolumn{4}{c}{\textbf{GJ~3090\,c}} \\
    $P$ [days] & $15.9510^{+0.0095}_{-0.0110}$ & $15.941^{+0.036}_{-0.041}$ & $15.9407^{+0.0057}_{-0.0060}$ \\
    $K$ [m/s] & $3.38^{+0.73}_{-0.70}$ & $4.02^{+0.55}_{-0.57}$ & $3.95^{+0.51}_{-0.52}$ \\
    $T_0$ [BJD-2457000] & $1350.60^{+0.84}_{-0.72}$ & $1351.4^{+5.5}_{-4.7}$ & $1350.83^{+0.63}_{-0.59}$ \\
    \midrule
    \multicolumn{4}{c}{\textbf{12.7-d Candidate}} \\
    $P$ [days] & $12.645^{+0.071}_{-0.015}$ & $12.728^{+0.499}_{-0.070}$ & $12.7138^{+0.0100}_{-0.0760}$ \\
    $K$ [m/s] & $2.90^{+0.56}_{-0.58}$ & $1.26^{+0.38}_{-0.36}$ & $1.76^{+0.42}_{-0.45}$ \\
    $T_0$ [BJD-2457000] & $1360.8^{+1.1}_{-1.0}$ & $1359.0^{+3.2}_{-3.4}$ & $1360.3^{+1.2}_{-1.3}$ \\
    \bottomrule
    \end{tabular}
\end{minipage}

\begin{figure}[htp]
    \centering
    \includegraphics[width=0.95\textwidth]{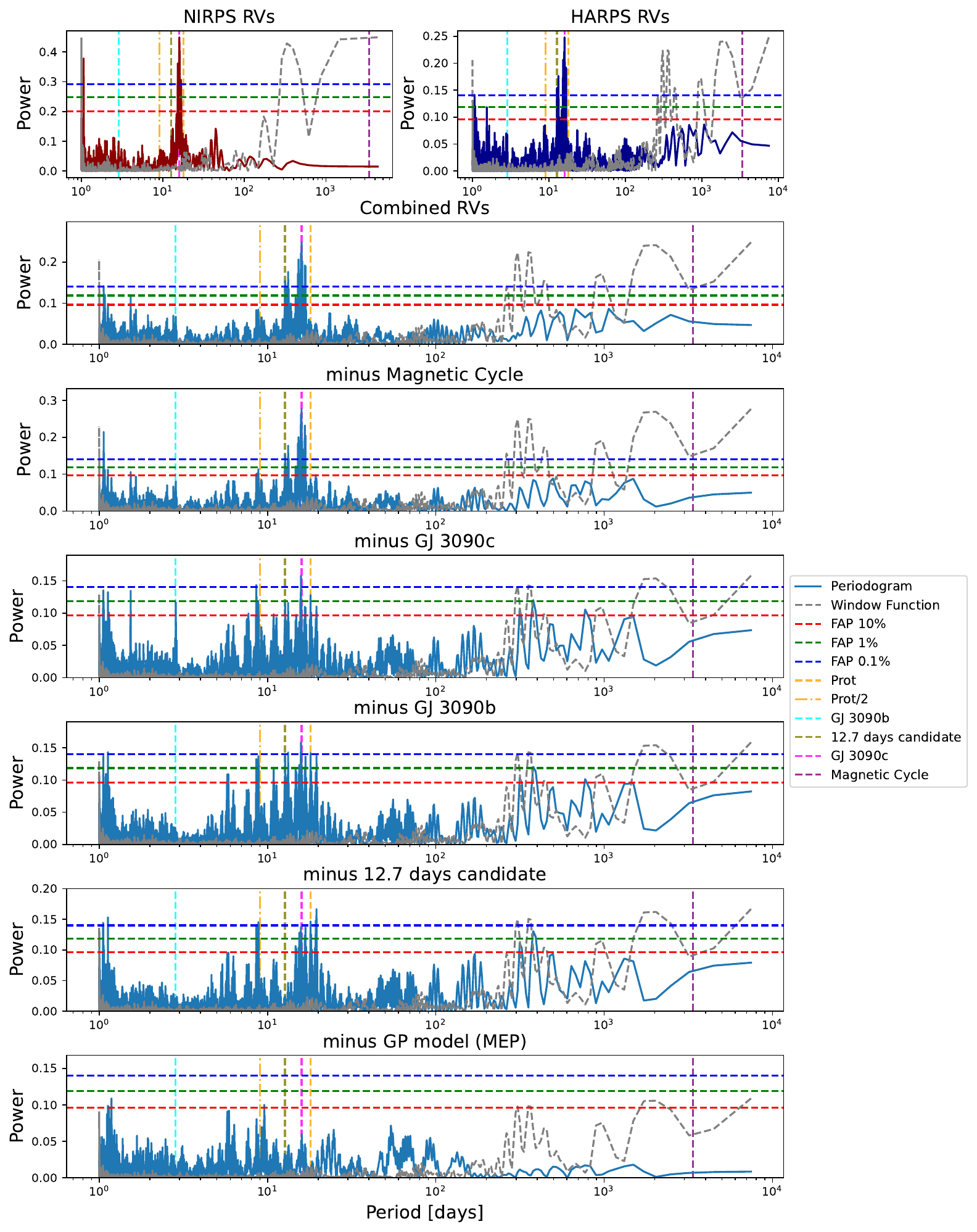}
    \caption{
        Periodograms of the RVs after having subtracted the best-fit offset of each instrument. The first panel shows the periodogram of the RVs of NIRPS and HARPS (HAM \& EGGS) separately. The second panel shows the periodogram of the combined RVs, with significant peaks marked for known planets and signals. The subsequent panels show the periodograms after sequentially removing the signals of the magnetic cycle, GJ~3090c, GJ~3090b, the 12.7-day candidate, and finally the GP model. Analytical false alarm probability (FAP) levels of 10\%, 1\%, and 0.1\% are indicated by dashed lines in red, green, and blue, respectively. The window function is shown as a grey dashed line.
    }
    \label{fig:periodograms}
\end{figure}

\begin{figure*}[h]
    \centering
    \includegraphics[width=1\textwidth]{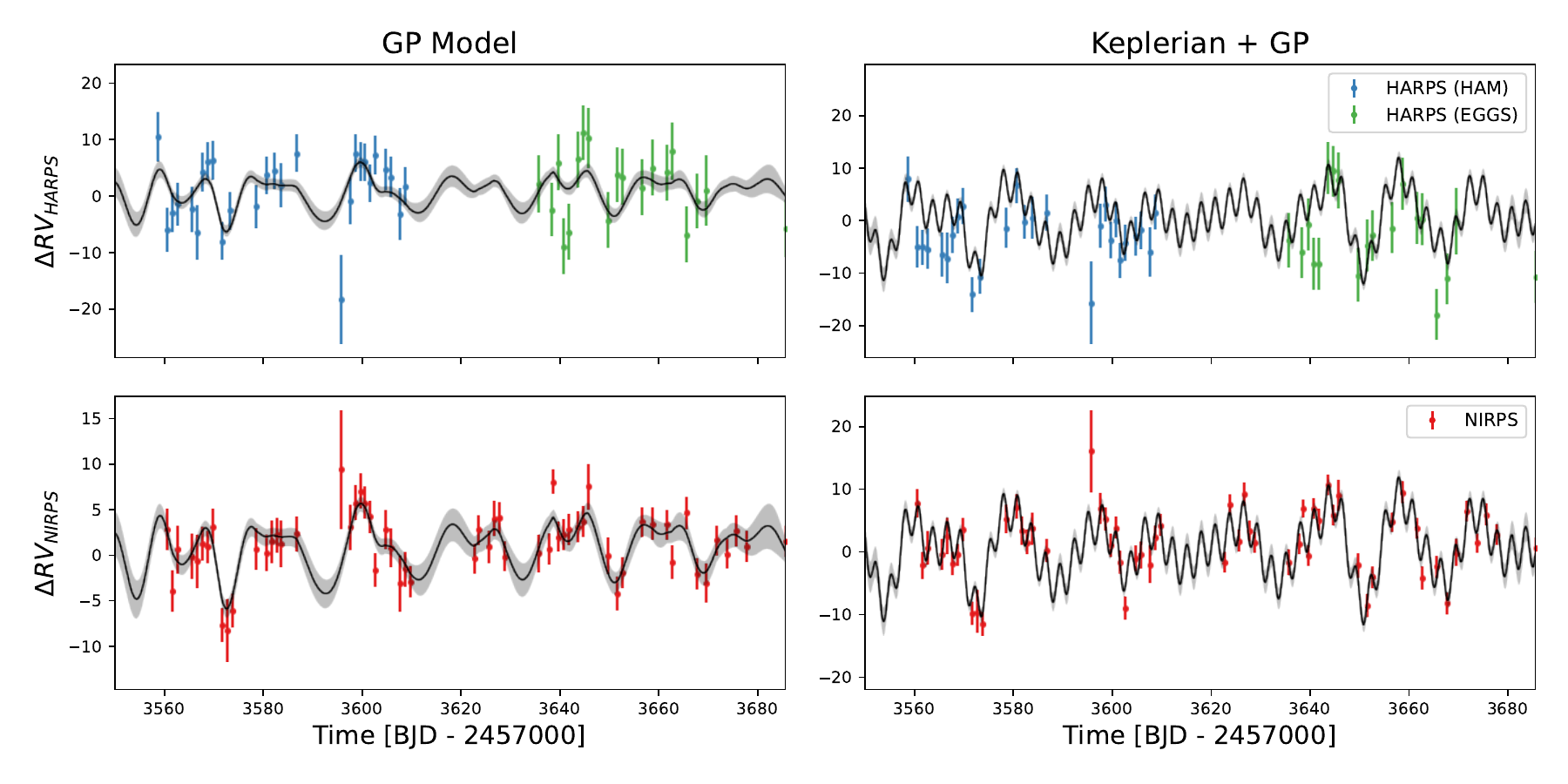} 

    \vspace{1cm}

    \includegraphics[width=1\textwidth]{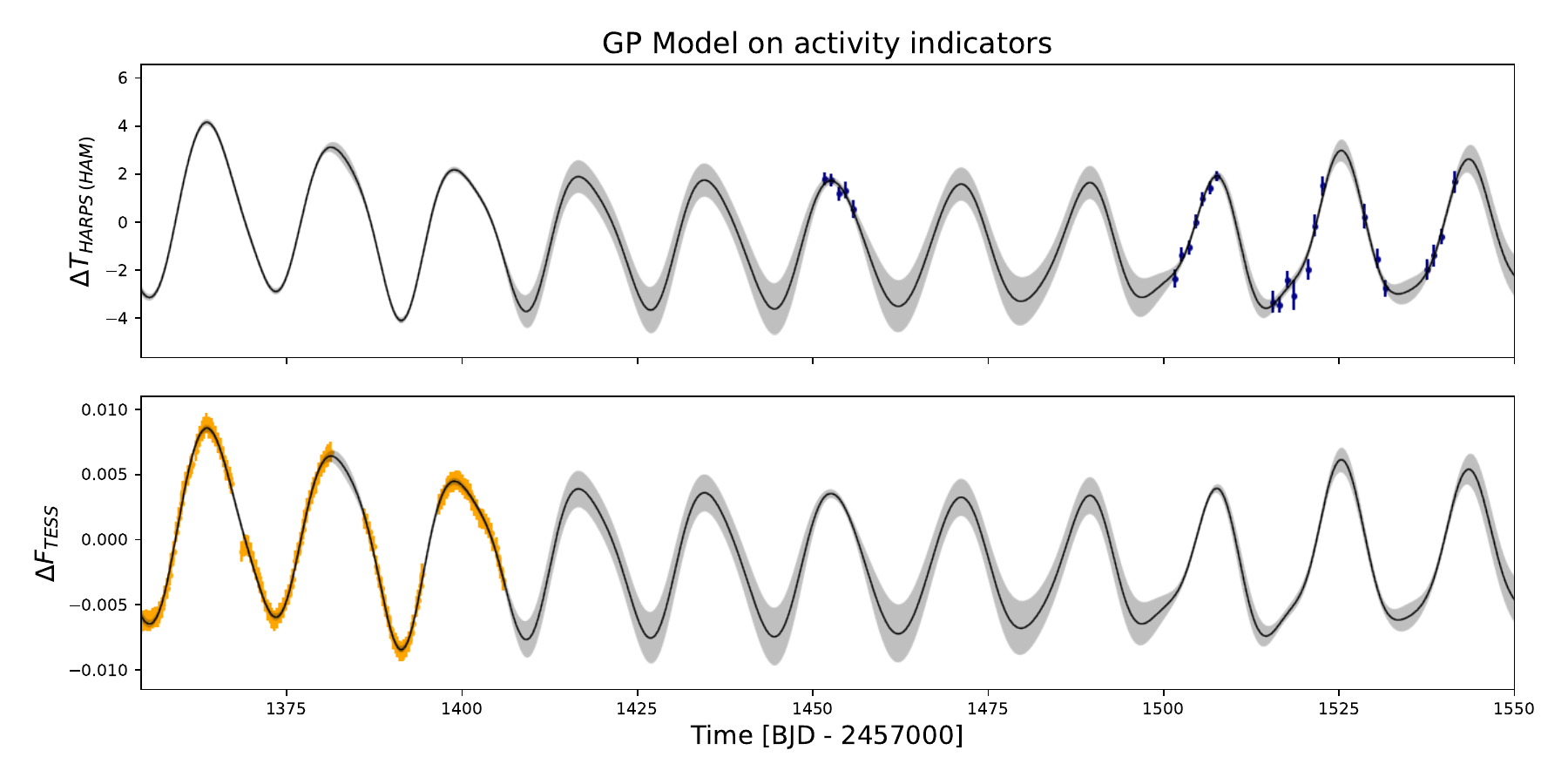} 

    \caption{\textbf{Top panel:} Subsets of the RV time series for HARPS and NIRPS. The left sub-panel shows the GP model fit to the RV data without the Keplerian signals, while the right sub-panel presents the combined Keplerian and GP model. \textbf{Bottom panel:} GP model applied to the activity indicators, specifically the $\Delta T$ and TESS data. The shaded regions represent the 1$\sigma$ confidence intervals of the GP model predictions. We note that the top and bottom panels do not show the same epochs of observation. The top panel shows 135 days of data before the last spectrum and the bottom panel shows up to 195 days after the beginning of first TESS sector.}
    \label{fig:timeseries}
\end{figure*}

\begin{figure*}[h]
    \centering
    \includegraphics[width=\textwidth]{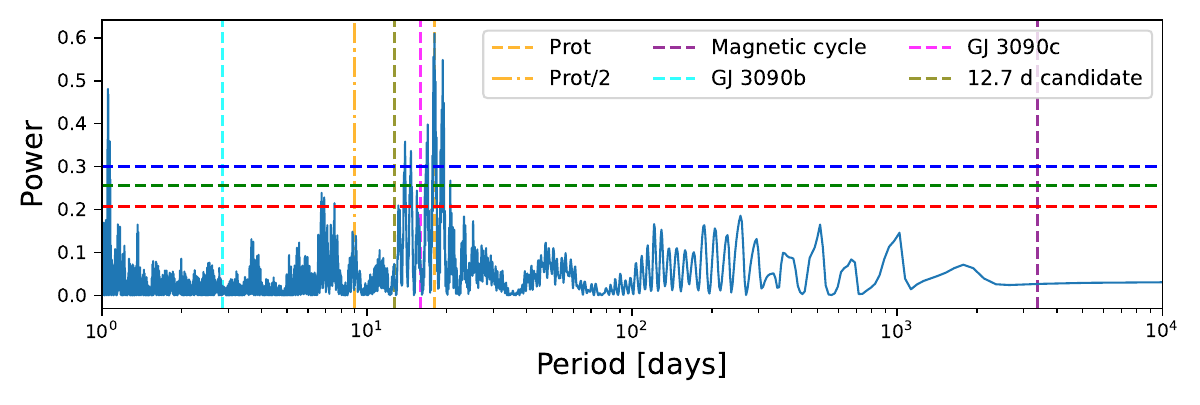}
    \caption{
        Lomb–Scargle periodogram of the HARPS (HAM) $\Delta T$ measurements.
        The solid blue line shows the power as a function of period, while the horizontal red, green, and blue dashed lines mark the 1\%, 0.1\%, and 0.01\% analytical false alarm probability thresholds respectively. There are no significant peaks around the planets' periods. There is a forest of statistically significant peaks between 13 and 20 days which shows a sign of potential differential rotation.
    }
    \label{fig:HARPS_HAM_dTemp_LS}
\end{figure*}

\begin{figure*}[h]
    \centering
    \includegraphics[width=\textwidth]{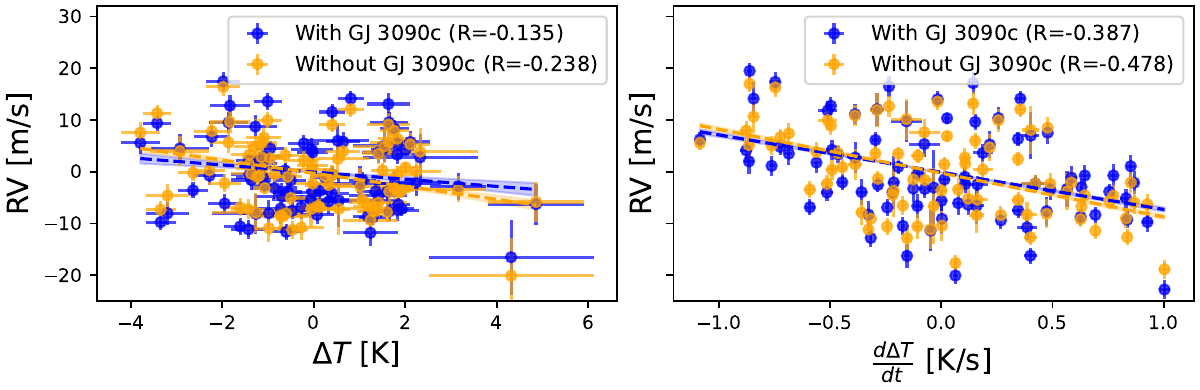}
    \caption{Correlations between HARPS (HAM) RVs and its $\Delta T$ indicator (left) and with the $d \Delta T/dt$ (right). The blue data points show the RVs that include the 16-day signal and the orange data points show the RVs after removing the best-fit Keplerian model of the 16-day signal. A linear function is fitted in each dataset. First-order correlation (top): $R=-0.135$ to $R=-0.238$ and $N_{\sigma} = 2.71$ to $N_{\sigma} = 6.73$. Second-order correlation (bottom): $R=-0.386$ to $R=-0.479$ and $N_{\sigma} = 17.2$ to $N_{\sigma} = 20.6$
    }
    \label{fig:correlation_rv_dtemp}
\end{figure*}

\begin{figure*}[h]
    \centering
    \includegraphics[width=\textwidth]{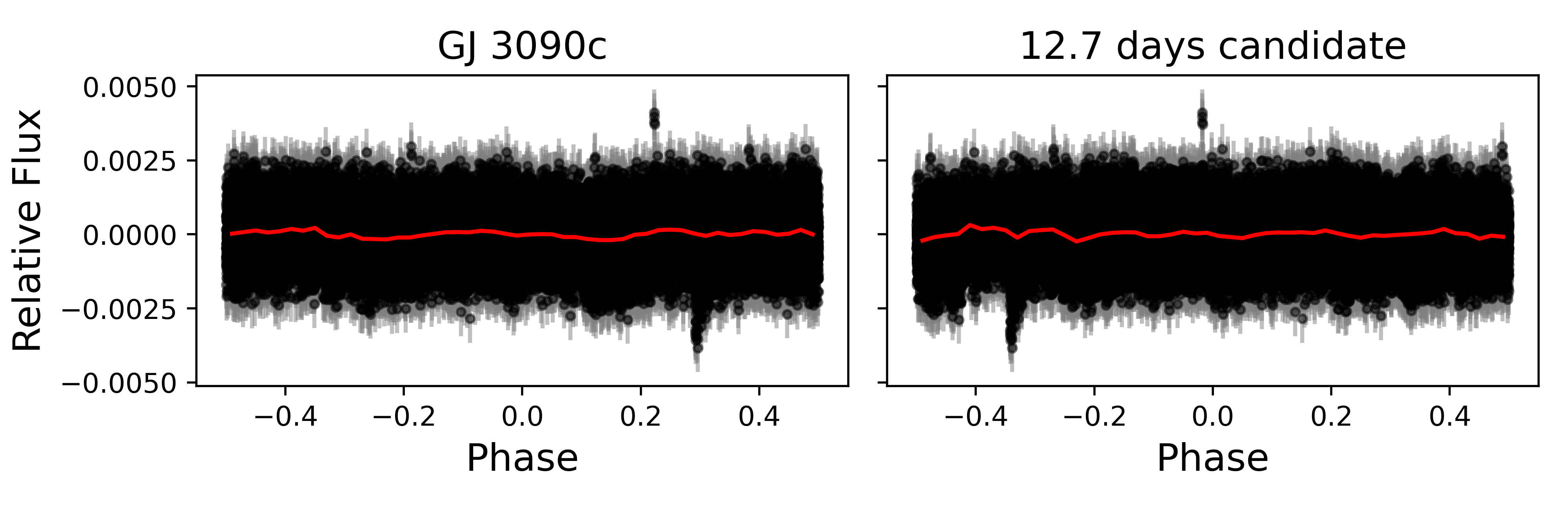}
    \caption{Phase-folded light curves for the two outer planets. The left panel shows the light curve for GJ 3090\,c, while the right panel corresponds to the 12.7-day candidate. The data points represent the TESS photometry corrected with the activity model. The red line is a binned version of the data ($50$ points per bin) so that we see the flux variations more easily. No transit is observed in either case.}
    \label{fig:phase_folded_light_curves}
\end{figure*}

\begin{figure*}[ht!]
    \centering
    \includegraphics[width=0.85\textwidth]{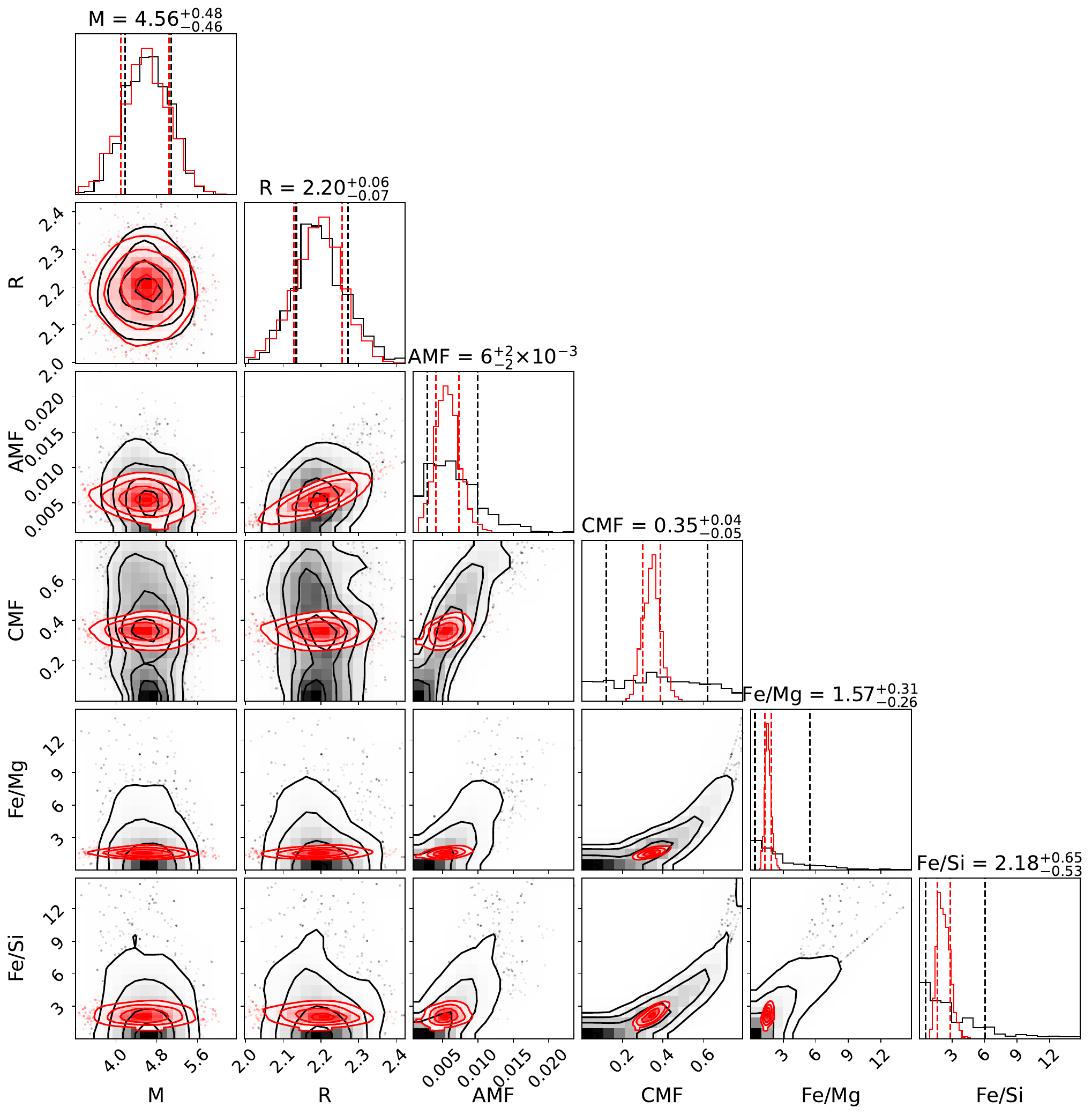}
    \caption{
    Corner plot for the H$_2$/He scenario of GJ~3090~b showing the posterior distributions for the planet mass ($M$), radius ($R$), atmospheric mass fraction (AMF), core mass fraction (CMF), and refractory element ratios (Fe/Mg and Fe/Si). 
    The red contours correspond to the case with stellar-informed priors on the refractory abundances, while the black contours correspond to the case without such priors. 
    The inclusion of stellar priors significantly reduces the spread in CMF and AMF values by constraining the rocky interior composition.
    }
    \label{fig:corner_H2He_priors}
\end{figure*}

\begin{figure*}[ht!]
    \centering
    \includegraphics[width=0.85\textwidth]{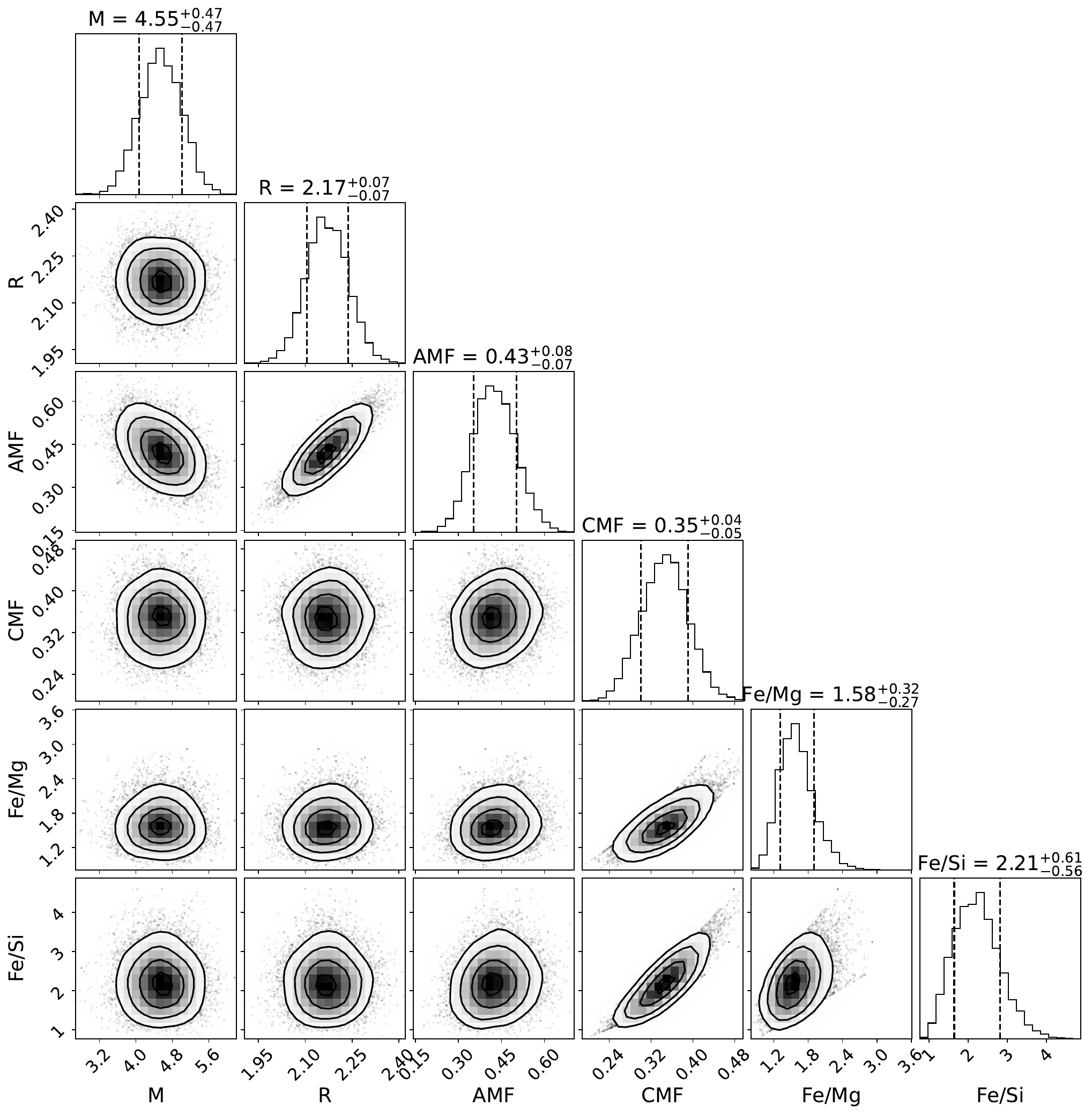}
    \caption{
       Corner plot for the pure H$_2$O atmosphere scenario of GJ~3090~b showing the posterior distributions for the planet mass ($M$), radius ($R$), water mass fraction (WMF), core mass fraction (CMF), and refractory element ratios (Fe/Mg and Fe/Si). 
    }
    \label{fig:corner_H2O_priors}
\end{figure*}

\clearpage

\begin{table*}[ht!]
\renewcommand{\arraystretch}{1.15}
\small
\centering
\caption{Priors and posteriors of the transit fit}
\begin{tabular}{llccc}
\toprule
\textbf{Parameter} & \textbf{Description} & \textbf{Prior} & \textbf{Posterior} & \textbf{Units} \\
\midrule
\multicolumn{5}{c}{\textbf{Keplerian Parameters}} \\
\midrule
\textbf{GJ~3090\,b} & & & & \\
$P$ & Orbital period & $\mathcal{N}(2.853, 0.010)$ & $2.85310198^{+0.00000080}_{-0.00000085}$ & days \\
$t_0$ & Time of conjunction & $\mathcal{N}(1356.153, 0.010)$ & $1356.15312^{+0.00027}_{-0.00026}$ & BJD$-2457000$ \\
$r_1$ & -- & $\mathcal{U}(0,1)$ & $0.798\pm0.014$ & -- \\ 
$r_2 = p$& Scaled planetary radius $R_p/R_\star$ & $\mathcal{U}(0,1)$ &  $0.03853^{+0.00050}_{-0.00047}$& -- \\ 
$b$& Impact parameter & -- & $0.698^{+0.020}_{-0.021}$ & -- \\ 
$i$& Inclination & -- & $86.93^{+0.16}_{-0.18}$ & deg \\ 
$e$ & Eccentricity & -- & 0 (fixed) & -- \\
$\omega$ & argument of periastron & -- & 90 (fixed) & deg \\
\midrule
\multicolumn{5}{c}{\textbf{Stellar Parameters}} \\
\midrule
$q_1$ & Limb darkening coeff & $\mathcal{N}(0.8453, 0.022)$ & $0.848^{+0.021}_{-0.022}$ &--  \\
$q_2$ & Limb darkening coeff & $\mathcal{N}(0.4177, 0.021)$ & $0.417^{+0.019}_{-0.020}$ & -- \\
$\rho$ & Stellar density & $\mathcal{N}(5205,428)$ & $5178^{+404}_{-407}$ & $g/cm^{-3}$ \\
\midrule
\multicolumn{5}{c}{\textbf{Gaussian Process Parameters}} \\
\midrule
$\rho_{GP,\mathrm{TESS}_{S2}}$ & GP Characteristic timescale & log$\mathcal{U}(0.001,50.0)$ & $0.199^{+0.018}_{-0.016}$ & days \\
$\sigma_{GP,\mathrm{TESS}_{S2}}$&  GP Amplitude & log$\mathcal{U}(10^{-6},50.0)$ & $5.03^{+0.35}_{-0.32}\times10^{-4}$ & -- \\
$\rho_{GP,\mathrm{TESS}_{S3}}$ & GP Characteristic timescale & log$\mathcal{U}(0.001,50.0)$ & $0.333^{+0.078}_{-0.052}$ & days \\
$\sigma_{GP,\mathrm{TESS}_{S3}}$ & GP Amplitude & log$\mathcal{U}(10^{-6},50.0)$ & $6.69^{+0.91}_{-0.70}\times10^{-4}$ & -- \\
$\rho_{GP,\mathrm{TESS}_{S29}}$ & GP Characteristic timescale & log$\mathcal{U}(0.001,50.0)$ & $0.435^{+0.101}_{-0.077}$ & days \\
$\sigma_{GP,\mathrm{TESS}_{S29}}$ & GP Amplitude & log$\mathcal{U}(10^{-6},50.0)$ & $4.31^{+0.61}_{-0.46}\times10^{-4}$ & -- \\
$\rho_{GP,\mathrm{TESS}_{S69}}$ & GP Characteristic timescale & log$\mathcal{U}(0.001,50.0)$ & $380^{+0.048}_{-0.041}$ & days \\
$\sigma_{GP,\mathrm{TESS}_{S69}}$ & GP Amplitude & log$\mathcal{U}(10^{-6},50.0)$ & $8.54^{+1.09}_{-0.83}\times10^{-4}$ & -- \\

\midrule
\multicolumn{5}{c}{\textbf{Instrument Offsets and Jitters}} \\
\midrule
$M_{\mathrm{TESS}_{S2}}$ & Offset relative flux & $\mathcal{N}(0.0, 0.03)$ & $-0.3^{+6.7}_{-6.3}\times10^{-5}$ & -- \\
$\sigma_{w,\mathrm{TESS}_{S2}}$ & Jitter & log$\mathcal{U}(0.01,300.0)$ & $0.7^{+8.5}_{-0.7}$ & ppm \\
$M_{\mathrm{TESS}_{S3}}$ & Offset relative flux & $\mathcal{N}(0.0, 0.03)$ & $-0.8^{+13}_{-14}\times10^{-5}$ & -- \\
$\sigma_{w,\mathrm{TESS}_{S3}}$ & Jitter & log$\mathcal{U}(0.01,300.0)$ & $0.6^{+9.5}_{-0.6}$ & ppm \\
$M_{\mathrm{TESS}_{S29}}$ & Offset relative flux & $\mathcal{N}(0.0, 0.03)$ & $-0.9^{+10}_{-9}\times10^{-5}$ & -- \\
$\sigma_{w,\mathrm{TESS}_{S29}}$ & Jitter & log$\mathcal{U}(0.01,300.0)$ & $179^{+24}_{-31}$ & ppm \\
$M_{\mathrm{TESS}_{S69}}$ & Offset relative flux & $\mathcal{N}(0.0, 0.03)$ & $-15\pm17\times10^{-5}$ & -- \\
$\sigma_{w,\mathrm{TESS}_{S69}}$ & Jitter & log$\mathcal{U}(0.01,300.0)$ & $2.1^{+41.8}_{-2.1}$ & ppm \\

\bottomrule
\end{tabular}
\label{tab:priors_posteriors_tess_fit}
\end{table*}

\begin{table*}[ht!]
\renewcommand{\arraystretch}{1.15}
\small
\centering
\caption{Priors and posteriors of the best-fit RV model (b4)}
\begin{tabular}{llccc}
\toprule
\textbf{Parameter} & \textbf{Description} & \textbf{Prior} & \textbf{Posterior} & \textbf{Units} \\
\midrule
\multicolumn{5}{c}{\textbf{Keplerian Parameters}} \\
\midrule
\textbf{GJ~3090\,b} & & & & \\
$P$ & Orbital period & $\mathcal{N}(2.85310198 \pm 8.2 \times 10^{-7})$ & $2.85310195 \pm 8.0 \times 10^{-7}$ & days \\
$K$ & RV semi-amplitude & $\mathcal{U}(0.0, 10.0)$ & $3.15 \pm 0.32$ & m/s \\
$t_0$ & Time of conjunction & $\mathcal{N}(1356.153115 \pm 0.00027)$ & $1356.15322 \pm 0.00023$ & BJD$-2457000$ \\
$e$ & Eccentricity & -- & 0 (fixed) & -- \\
$\omega$ & argument of periastron & -- & 90 (fixed) & deg \\
\textbf{GJ~3090\,c} & & & & \\
$P$ & Orbital period & $\mathcal{U}(14.0, 20.0)$ & $15.9407 \pm 0.0058$ & days \\
$K$ & RV semi-amplitude & $\mathcal{U}(0.0, 10.0)$ & $3.95 \pm 0.51$ & m/s \\
$t_0$ & Time of conjunction & $\mathcal{U}(1340.0, 1360.0)$ & $1350.83 \pm 0.61$ & BJD$-2457000$ \\
$e$ & Eccentricity & -- & 0 (fixed) & -- \\
$\omega$ & argument of periastron & -- & 90 (fixed) & deg \\
\textbf{12.7d Candidate} & & & & \\
$P$ & Orbital period & $\mathcal{U}(10.0, 14.0)$ & $12.7138_{-0.0760}^{+0.0100}$ & days \\
$K$ & RV semi-amplitude & $\mathcal{U}(0.0, 10.0)$ & $1.76 \pm 0.44$ & m/s \\
$t_0$ & Time of conjunction & $\mathcal{U}(1353.0, 1367.0)$ & $1360.3_{-1.3}^{+1.2}$ & BJD$-2457000$ \\
$e$ & Eccentricity & -- & 0 (fixed) & -- \\
$\omega$ & argument of periastron & -- & 90 (fixed) & deg \\
\midrule
\multicolumn{5}{c}{\textbf{Gaussian Process Parameters}} \\
\midrule
$P_{\rm rot}$ & Stellar rotation period & $\mathcal{U}(14, 22)$ & $17.95 \pm 0.15$ & days \\
$\rho$ & Evolution timescale & $\mathcal{L}(9, 360)$ & $126_{-32}^{+41}$ & days \\
$\eta$ & Periodicity scale & $\mathcal{L}(0.1, 10)$ & $0.547 \pm 0.083$ & -- \\
$\alpha_0$ & GP amp. (HARPS) & $\mathcal{U}(-50, 50)$ & $-1.54_{-0.74}^{+0.67}$ & m/s \\
$\beta_0$ & GP' amp. (HARPS) & $\mathcal{U}(-50, 50)$ & $-17.0_{-3.4}^{+2.8}$ &  m/s \\
$\alpha_1$ & GP amp. (NIRPS) & $\mathcal{U}(-50, 50)$ & $-1.9_{-1.2}^{+1.3}$ &  m/s \\
$\beta_1$ & GP' amp. (NIRPS) & $\mathcal{U}(-50, 50)$ & $-15.8_{-4.1}^{+3.6}$ &  m/s \\
$\alpha_2$ & GP amp. ($\Delta T$) & $\mathcal{L}(10^{-7}, 5)$ & $2.15_{-0.30}^{+0.37}$ & K \\
$\alpha_3$ & GP amp. (TESS) & $\mathcal{L}(10^{-7}, 0.01)$ & $0.00444 \pm 0.00076$ & Norm. flux \\
\midrule
\multicolumn{5}{c}{\textbf{Magnetic Cycle Parameters}} \\
\midrule
$P_{\rm cyc}$ & Cycle period & $\mathcal{N}(3370, 700)$ & $3285_{-526}^{+637}$ & days \\
$\tau_{\rm H}^{\rm cyc}$ & Cycle $t_0$ (HARPS) & $\mathcal{U}(-1144, 3856)$ & $1256_{-278}^{+190}$ & BJD$-2457000$ \\
$K_{\rm H}^{\rm cyc}$ & Cycle amp. (HARPS) & $\mathcal{L}(10^{-7}, 20)$ & $4.14_{-0.90}^{+1.07}$ & m/s \\
$\tau_{\rm N}^{\rm cyc}$ & Cycle $t_0$ (NIRPS) & $\mathcal{U}(-1144, 3856)$ & $730_{-1256}^{+1753}$ & BJD$-2457000$ \\
$K_{\rm N}^{\rm cyc}$ & Cycle amp. (NIRPS) & $\mathcal{L}(10^{-7}, 20)$ & $0.000060_{-0.000060}^{+0.059583}$ & m/s \\
$\tau_{\Delta T}^{\rm cyc}$ & Cycle $t_0$ ($\Delta T$) & $\mathcal{U}(-1144, 3856)$ & $1151_{-1520}^{+1630}$ & BJD$-2457000$ \\
$K_{\Delta T}^{\rm cyc}$ & Cycle amp. ($\Delta T$) & $\mathcal{L}(10^{-7}, 5)$ & $0.000075_{-0.000075}^{+0.052736}$ & K \\
\midrule
\multicolumn{5}{c}{\textbf{Instrument Offsets and Jitters}} \\
\midrule
$\gamma_{\rm HAM}$ & HARPS-HAM offset & $\mathcal{N}(17233.64, 7.74)$ & $17233.59 \pm 0.89$ & m/s \\
$\gamma_{\rm EGGS}$ & HARPS-EGGS offset & $\mathcal{N}(17187.33, 6.36)$ & $17189.1 \pm 1.4$ & m/s \\
$\gamma_{\rm NIRPS}$ & NIRPS offset & $\mathcal{N}(17181.78, 5.38)$ & $17179.93_{-0.91}^{+0.73}$ & m/s \\
$\gamma_{\Delta T}$ & $\Delta T$ offset & $\mathcal{N}(-0.06, 1.72)$ & $0.04 \pm 0.58$ & K \\
$\gamma_{\rm TESS}$ & TESS offset & $\mathcal{N}(1.00003, 0.0041)$ & $1.0001 \pm 0.0014$ & Norm. flux \\
$\sigma_{\rm jit, HAM}$ & HARPS-HAM jitter & $\mathcal{L}(10^{-5}, 77.41)$ & $2.97 \pm 0.47$ & m/s \\
$\sigma_{\rm jit, EGGS}$ & HARPS-EGGS jitter & $\mathcal{L}(10^{-5}, 63.56)$ & $4.71_{-0.68}^{+0.78}$ & m/s \\
$\sigma_{\rm jit, NIRPS}$ & NIRPS jitter & $\mathcal{L}(10^{-5}, 53.80)$ & $0.0052_{-0.0052}^{+0.9705}$ & m/s \\
$\sigma_{\rm jit, \Delta T}$ & $\Delta T$ jitter & $\mathcal{L}(10^{-5}, 17.16)$ & $0.0021 \pm 0.0254$ & K \\
$\sigma_{\rm jit, TESS}$ & TESS jitter & $\mathcal{L}(10^{-5}, 0.0409)$ & $0.0000215 \pm 0.0000160$ & Norm. flux \\
\bottomrule
\end{tabular}
\label{tab:priors_posteriors}
\end{table*}

\end{document}